\documentclass[11pt]{article}
\usepackage{graphicx}
\usepackage{jheppub}
\usepackage{comment}
\usepackage{xcolor}
\usepackage{cancel}
\usepackage{threeparttable}
\newcommand{\df}{\mathrm d}
\newcommand{\sub}[1]{_{\mathrm{#1}}}

\newcommand{\vderiv}[2]{\frac{\delta #1}{\delta #2}}

\newcommand{\MM}{{\mathcal M}}
\newcommand{\pMM}{{\partial\mathcal M}}
\DeclareMathOperator{\sign}{sign}

\DeclareMathOperator{\sech}{sech}

\definecolor{rcolor}{rgb}{0.9,0.1,0.1}

\definecolor{mcolor}{rgb}{0.8,0.1,0.6}

\definecolor{gcolor}{RGB}{0,100,0}

\definecolor{gcolor}{RGB}{0,100,0}

\definecolor{bcolor}{RGB}{0,0,200}

\title{Scalar/axion Wormholes and Boundary Conditions in gravitational path integrals}
\author[a]{Jason Hallsten,} \emailAdd{jhalsten@ethz.ch}
\author[b]{Jesse Held,} \emailAdd{jheld@ucsb.edu}
\author[c]{Zhenyu Heng,} \emailAdd{zhenyuheng@umail.ucsb.edu}
\author[d]{Donald Marolf,} \emailAdd{marolf@ucsb.edu}
\author[e,f]{Zhencheng Wang} \emailAdd{zhencheng.wang@colorado.edu}

\affiliation[a]{Department of Physics, ETH Z\"urich, 8093 Z\"urich, Switzerland}
\affiliation[b]{Northampton, MA, USA}
\affiliation[c]{Department of Physics, Huazhong University of Science and Technology, Wuhan, Hubei 430074, China}
\affiliation[d]{Department of Physics, University of California, Santa Barbara, CA 93106, USA}
\affiliation[e]{Department of Physics, University of Illinois Urbana-Champaign, Urbana, IL 61801, USA}
\affiliation[f]{Department of Physics, University of Colorado, Boulder, CO 80309, USA}

\abstract{Recent work noted that whether or not a given class of spacetime wormholes contributes to the gravitational path integral can depend strongly on the choice of boundary conditions.  In particular, for asymptotically-AdS scalar/axion wormholes with spherical boundaries it was suggested that there are regimes where the same wormhole contributes to a Euclidean low-energy effective field theory computation of the partition function $Z_N$ defined by imposing Neumann boundary conditions on the scalar/axion field, while it fails to contribute to 
a corresponding partition function $Z_D$ defined by imposing Dirichlet boundary conditions on this field.   Here we investigate toy models of related wormholes associated with JT gravity coupled to a scalar field, asymptotically-AdS gravity coupled to an axion/scalar with toroidal boundary conditions, and asymptotically flat gravity coupled to an axion/scalar.  In all cases we compare $Z_D$ and $Z_N$ as computed in a truncated low-energy effective theory by an appropriate integral over constrained wormholes.  In each case we find that, even though $Z_D$ and $Z_N$ are related by an integral transform,  the two generally receive contributions from differing sets of saddles.  Along the way, we uncover a variety of additional interesting subtleties associated with semiclassical contributions to our $Z_D$ and $Z_N$.
}

\begin{document}
\maketitle
\section{Introduction}

Euclidean spacetime wormhole saddles for the gravitational path integral continue to be of great interest in connection with the AdS/CFT factorization problem \cite{Maldacena:2004rf}.  We focus below on smooth wormholes such as the original axion wormholes of \cite{Giddings:Original_Axion_Wormhole}, their various axion and scalar cousins \cite{Halliwell:1989pu,aguilar:Axion_dS_Wormholes,Arkani-Hamed:2007cpn,Garcia-Garcia:2020ttf,Garcia-Garcia:2022zmo,andriolo:Axion_wormholes_massive_dilaton}, the various constructions of \cite{Maldacena:2004rf}, and wormholes supported by gradients of matter fields \cite{Marolf:2021kjc}, though other interesting wormholes include the double cone \cite{Saad:2018bqo} and wormholes supported by Euclidean particle worldlines \cite{Chandra:2022bqq,Balasubramanian:2022gmo,Wang:2025bcx}.

If we assume the wormhole to have a $U(1)$ isometry, or a larger symmetry group that includes a $U(1)$, such wormholes necessarily require imaginary (or complex) sources.  This can be seen from the fact that Wick rotation to Lorentz signature yields a time-independent wormhole which is thus traversable.  But classical traversable wormholes are forbidden if all matter fields satisfy the null energy condition (NEC) \cite{Friedman:1993ty}.  Imaginary or complex sources are then required if one is to use standard matter fields which satisfy NEC for real field values.  It is an interesting question to what extent one might avoid the use of imaginary or complex sources in less symmetric contexts\footnote{The hyperbolic wormholes of \cite{Maldacena:2004rf} have no such symmetries, though even with mass-deformations such wormholes are known only when the deformation is so small that the theory remains unstable \cite{Marolf:2021kjc}.}.

The above imaginary and complex sources raise interesting questions about the consistency of the underlying microscopic theory at such couplings \cite{Maldacena:2026jqd}.  In addition,  and at a more practical level, they raise questions about how one should determine whether such wormholes in fact contribute to a given physical computation even at the level of the low-energy effective theory.  Studies of the latter question were recently performed in \cite{AxionWormholes,Held:2026bbo} by considering an integral over constrained saddles for the topology-changing version of the Lorentz-signature gravitational path integral described in \cite{Marolf:Thermodynamics} (building on \cite{Louko:1995jw,Dong:2016hjy,Colin-Ellerin:2020mva,Marolf:2020rpm,Colin-Ellerin:2021jev} as well as previous studies of constrained wormholes \cite{Cotler_2021:ConstrainedInstantons,Cotler:2021cqa}).  However, after performing a certain integral transform,  the actual computations for the uncharged static cases discussed in \cite{AxionWormholes,Held:2026bbo} reduced to an integral over constrained saddles that could also be motivated from purely Euclidean considerations.  We will thus not concern ourselves  here with any other microscopic consistency or general prescriptions for formulating gravitational path integrals. Instead, we will simply jump directly to the technical point that motivates the work below\footnote{We note, however, that like many earlier works \cite{Hartle:2020glw,Schleich:1987fm,Mazur:1989by,Giddings:1989ny,Giddings:1990yj,Marolf:1996gb,Dasgupta:2001ue,Ambjorn:2002gr,Feldbrugge:2017kzv,Feldbrugge:2017fcc,Feldbrugge:2017mbc,Brown:2017wpl}, \cite{Marolf:Thermodynamics} endorsed a Lorentzian prescription in order to avoid the well-known conformal factor problem \cite{Gibbons:1978ac} of Euclidean gravity which necessarily requires that any ``Euclidean'' approach be associated with a non-trivial contour prescription for integrating over an appropriate class of complexified ``Euclidean'' metrics. }.

In particular, \cite{AxionWormholes,Held:2026bbo} first imposed a constraint on the area $A_0$ of a certain cross-section of the wormhole that allowed the existence of constrained wormhole saddles with real matter fields. Here, following \cite{AxionWormholes} and \cite{Witten:2026twr}, though in contrast to most other references, the sign of the kinetic term for a real axion is taken to be the same as that for a real scalar even in Euclidean signature (so that the axion wormholes of e.g. \cite{Giddings:Original_Axion_Wormhole} require imaginary axions).  The relevant $A_0$ was sometimes called $L_0$ in low dimensions.  The constrained saddles were then used to give a semiclassical approximation to the fixed-$A_0$ sector of the wormhole path integral, after which the desired partition function was defined in terms of an  integral over positive real $A_0$ that was performed carefully.  Again, while these works started in Lorentz signature, at the appropriate stage their computations reduced to the study of real Euclidean-signature wormhole metrics supported by the fixed-$A_0$ constraint and by real values of the matter fields.

An interesting feature of this analysis was that, despite the results of various previous studies \cite{Loges:2022nuw,andriolo:Axion_wormholes_massive_dilaton,Loges:2023ypl,Hertog:Axion-Saxion_Wormholes_Stable,aguilar:Axion_dS_Wormholes,loveridge:Axion_Wormhole_Alternate_Topologies,Marolf:2025evo}
finding axion wormholes to be `stable,' ref.  \cite{AxionWormholes} found that the standard axion wormholes (with purely imaginary axions in our conventions) did not contribute to the path integral studied in their main text.\footnote{More precisely, this case corresponded to a Stokes' line on which the wormhole failed to contribute on one side.} As noted in \cite{AxionWormholes}, in addition to potential mathematical subtleties and the Lorentzian starting point of \cite{AxionWormholes}, this difference could be plausibly traced to differences in the choice of boundary conditions.  Appendix E of \cite{AxionWormholes} then found evidence that, while a given wormhole saddle will have some definite value at the boundary for both the axion field and its normal derivative, it can contribute to, and in fact can even dominate, the partition function $Z_N$ defined by the relevant Neumann boundary condition even when it fails to contribute to the partition function $Z_D$ defined by the Dirichlet boundary condition.

We thus investigate other related wormholes below, comparing their contributions to Dirichlet and Neumann partition function.  In particular, we study the scalar wormhole of \cite{Garcia-Garcia:2020ttf} in Jackiw-Teitelboim (JT) gravity with matter, asymptotically AdS axion wormholes with toroidal boundaries, and asymptotically flat axion wormholes \cite{Giddings:Original_Axion_Wormhole}.   In each of these models the couplings are sufficiently simple that there is in fact no difference between using scalars or axions. The choice of terminology is thus merely historical, or a matter of convenience, and we will use the terms scalar and axion interchangeably for these models. In particular, whether we refer to our matter field as an axion or a massless scalar, we will consider this field to take values in ${\mathbb R}$; i.e., we take the target space to be noncompact.

In contrast to \cite{AxionWormholes}, for reasons of simplicity and familiarity we work entirely in Euclidean signature.  However, in the first two cases above, the constrained wormholes over which we integrate feature a nowhere-vanishing hypersurface-orthogonal $U(1)$ Killing field along which one can easily Wick-rotate to translate between Euclidean and Lorentzian perspectives.  

As a result, for these cases we can easily compute the energy of our constrained wormholes as a function of $A_0$.  An important observation is that, for what we call real-valued Neumann boundary conditions (i.e., for real values of the scalar/axion derivative at the asymptotic boundary), this energy is unbounded below.  In particular, in the limit  $A_0 \rightarrow 0$ this energy becomes infinitely negative.\footnote{This is the limit where the wormhole pinches off, and so is naturally UV sensitive. It is therefore interesting to ask if this problem might be removed by UV effects, though such issues are beyond the scope of our analysis here.}  The usual notion of a Neumann partition function is thus ill-defined in this approach.  In particular, as noted above, it would naturally be defined by integrating $e^{-S_E}$ over $A_0 \in {\mathbb R}^+$, and this integral simply turns out to diverge at the endpoint $A_0=0$.
We emphasize that, due to the above-mentioned Killing field, arbitrarily negative energies must also arise in the Lorentzian approach of \cite{AxionWormholes} (which explicitly required a formulation in which the energy is bounded below).

On the other hand, our conventions for the definition of `real' Neumann boundary conditions are tied to our convention that the kinetic term of a Euclidean axion/scalar should have the standard (positive) sign.  It follows that  the smooth Euclidean wormholes in each of these models (e.g., from \cite{Giddings:Original_Axion_Wormhole}) occurs at what we call {\it imaginary} Neumann boundary conditions  (i.e., for imaginary values of the scalar/axion derivative at the asymptotic boundary).  Furthermore, for such imaginary boundary conditions, at least in the JT+scalar case our wormhole energies remain real and become bounded below, so that a Neumann partition function $Z_N$ can be well-defined.  

We will thus adopt the approach of first defining $Z_N$ for imaginary boundary conditions and then analytically continuing the result to real boundary conditions as desired.  
Nevertheless, the reader would be right to question the physics of an approach that begins with what would normally be called an unphysical boundary condition and then defines the supposedly physical regime by analytic continuation.  We therefore provide three remarks that address this concern.

The first remark is that further motivation for the start-with-imaginary boundary conditions approach to $Z_N$ comes from consideration of the Lorentz-signature theory.  In the Lorentzian case, Dirichlet and Neumann path integrals are related by a Fourier transform.  Fourier transforms converge (at least distributionally) on the space of tempered distributions, which even allows sub-exponential growth of the function to which the transform is applied.  However, Wick rotation to Euclidean signature introduces an additional factor of $i$ so that, with real boundary conditions, we instead have an exponential transform integrated over the entire real line, and which is thus much less likely to converge.  For analytic actions, choosing to start with imaginary Neumann boundary conditions in the Euclidean case can thus again be related to a Fourier transform of the Dirichlet partition function $Z_D$, preserving a natural relation between Dirichlet and Neumann quantities that is present in the Lorentzian theory.

Indeed, this argument in fact suggests an improvement that will prove necessary in two of the systems we study.  In the toroidal-AdS and asymptotically-flat cases, for imaginary Neumann boundary conditions our Neumann constrained-wormhole actions will turn out to have branch points on the positive real $A_0$-axis.  As a result, the above proposal of defining $Z_N$ for imaginary Neumann boundary conditions is well-defined only after giving a prescription for dealing with the various possible choices of branches.  When this arises we will therefore simply {\it define} the Neumann partition function $Z_N$ (at imaginary Neumann boundary conditions) as the appropriately-normalized Fourier transform of $Z_D$.  Note that, had we not made such a choice, or if the definitions of the Neumann and Dirichlet partition functions $Z_N,Z_D$ were not coordinated in some natural way, there would then be no reason at all to expect the same saddles to govern the asymptotics of $Z_D$ and $Z_N$.  The interesting fact is that we will find different sets of wormholes to contribute to $Z_N$ and $Z_D$ despite defining them to be related via a Fourier transform.

The second remark is that the
arbitrarily negative energies for real Neumann boundary conditions may in the end turn out to be a consequence of the particular constrained wormholes over which we choose to integrate.  In Euclidean signature, such a choice is tantamount to at least a partial specification of a prescription for dealing with the so-called conformal factor problem \cite{Gibbons:1978ac}, and it may simply be that this particular prescription is incompatible with real Neumann boundary conditions.  This, however, does not rule out the possibility that a different such prescription might in fact lead to a well-defined $Z_N$ for real boundary conditions.  Indeed, even in the Dirichlet case and in Lorentz signature, an early attempt at the analysis in \cite{AxionWormholes} that used a different set of constraints was abandoned precisely because it led to arbitrarily negative energies,\footnote{While there is no conformal factor problem in Lorentz signature, the oscillatory nature of the Lorentzian path integral can make it sensitive to the order in which the various integrations are performed.  Focusing on a family of constrained wormholes effectively saves the integrals over the constraint parameters for last, so changing the constraints also alters the order of integrations and can affect convergence.}  It may thus be that a different set of constraints may be more successful here as well, or that our analytically-continue-from-imaginary-values prescription can be justified by some more complete analysis.

However, our final remark on this issue is simply that, as noted above, the microscopic physics of these models is not the primary concern of this work.  Instead, we will treat simple truncations of our systems as toy models to explore the possible relations between contributions of wormholes to Dirichlet and Neumann partition functions.  A prescription that defines $Z_N$ in terms of the Fourier transform of $Z_D$ appears to be quite sufficient for this purpose.  

Interestingly, as emphasized in \cite{AxionWormholes},  imposing Neumann boundary conditions on massless axions or scalars in asymptotically AdS settings is generally expected to lead to ghosts or other pathologies \cite{andrade:BeyondUnitarityBound}.  However, since such issues are associated with boundary kinetic terms, one expects to be able to fix Neumann data for the {\it average} of the scalar/axion over the boundary (i.e., for the zero-mode) so long as one fixes Dirichlet data for all modes with non-trivial boundary momentum\footnote{We thank Maciej Kolanowski for emphasizing this point to us.}.  Since in practice we assume homogeneity over the boundary, it is easy for us to adopt this point of view. As a result, the issues with negative energies discussed above appear to have no relation to such ghosts.

We are now ready to analyze the various systems of interest.
We address
the scalar wormhole of \cite{Garcia-Garcia:2020ttf} in Jackiw-Teitelboim (JT) gravity with matter, asymptotically AdS axion wormholes with toroidal boundaries, and asymptotically flat axion wormholes in turn in sections \ref{G3wormhole}, \ref{sec:tori}, and \ref{sec:asFlat}.  We then conclude with some final discussion in section \ref{sec:Disc}.

\section{JT gravity with a massless scalar}
\label{G3wormhole}

Following \cite{Garcia-Garcia:2020ttf}, we consider Jackiw-Teitelboim (JT) gravity, which is minimally coupled to a massless scalar field $\chi$ in an asymptotically AdS$_2$ spacetime.  The action thus takes the form $S = S_{JT} + S_{matter}$. In Euclidean signature we have 
\begin{equation}
    S_{JT} = - \frac{S_0}{2\pi}\left[ \frac{1}{2}\int_{\mathcal{M}} d^2x \sqrt{g}R + \int_{\partial \mathcal{M}} dt \sqrt{h}K \right] - \frac{1}{2}\int_{\mathcal{M}} d^2x \sqrt{g}\Phi(R+2) - \int_{\partial \mathcal{M}} dt \sqrt{h}{\Phi}(K-1),
\end{equation}
and
\begin{equation}
    S_{matter} = \frac{1}{2}\int_{\mathcal{M}} d^2x \sqrt{g} \partial_\mu \chi \partial^\mu \chi + S_{bdy, \chi}. \label{eq:SmatterJT}
\end{equation}
The relevant semiclassical limit will be defined by taking the boundary values of the dilaton $\Phi$ and the scalar $\chi$ to be large as explained below.  We also note that, since all of our calculations are on cylinders, the topological term multiplying $S_0$ simply vanishes.  We will thus not consider it further below.

We wish to compute the associated connected two-boundary partition function defined by a pair of asymptotically AdS$_2$ circular boundaries, both corresponding to boundary circles of the same size $\beta$, using either Dirichlet or Neumann boundary conditions on the scalar.  We refer to one boundary as `left' and the other as `right.' Boundary conditions on the dilaton are fixed by requiring that the length $\beta$ be defined using a Fefferman-Graham coordinate $z$ in which the asymptotic dilaton profile is $\Phi = \phi_b z^{-1} + O(z)$. Equivalently, we may say that we
evaluate the theory at a regularized asymptotic boundary with an induced metric satisfying $\sqrt{h} = 1/\epsilon$ and at which $\Phi = \phi_b/\epsilon$, after which we take the limit $\epsilon \rightarrow 0$. For Dirichlet boundary conditions, we will fix the boundary values of the scalar $\chi$ to be $\chi_R=+k/2$ and $\chi_L=-k/2$, so that $k:=\chi_R-\chi_L$ and for simplicity we have assumed $\chi$ to be antisymmetric. For  Neumann boundary conditions, it is instead convenient to fix the flux density
\begin{equation}
    q_{R,L}:=\frac{1}{\beta}\int_0^\beta dt_E\sqrt{h}\,n_{R,L}^\mu\partial_\mu\chi,
    \qquad q_R=-q_L=q=C_0
\end{equation}
on each boundary, where $n_R,n_L$ are the outward-directed normals and
where $C_0:=\sqrt f\,\chi'$ is the conserved flux density. Since the flux is conserved, wormhole solutions can exist only for $q_R = -q_L =  q$.  The parameters $\phi_b, k, \chi_R, \chi_L, q$ should all be taken to be large in the semiclassical limit.

We will first analyze the desired partition function for real values of $k, q$, but we will find the Neumann partition function to be divergent.  This will then require us to instead define $Z_N$ at imaginary $q$. The Dirichlet case allows the choice
$S_{bdy, \chi}=0$, while the Neumann case requires the non-trivial boundary term
\begin{equation}
    S_{bdy, \chi} = -\int_{\partial \mathcal{M}} dt \sqrt{h} \chi n^\mu \partial_\mu \chi.
\end{equation}

As in \cite{Garcia-Garcia:2020ttf}, there will be no wormhole solutions with real $k, q$.  We will thus follow \cite{AxionWormholes,Held:2026bbo} in imposing constraints, finding constrained saddles for which $e^{-S_E}$ is expected to approximate the corresponding constrained partition function, and then finally integrating over the constrained parameters to compute the desired partition functions.  The Dirichlet computation was already performed in \cite{Held:2026bbo} by constraining the $\chi=0$ surface to have length $A_0$.  As described in \cite{Held:2026bbo}, the constraint allows a discontinuity in the derivative of the dilaton but requires continuity of the metric, dilaton, and extrinsic curvature.

However, directly implementing this constraint is more difficult for Neumann boundary conditions where the boundary conditions preserve the shift symmetry $\chi \to \chi + C$.  As a result of this symmetry,  configurations satisfying the specified boundary conditions may fail to contain any surface on which $\chi$ vanishes. For this case we therefore instead choose to fix the length $A_0$ of the surface with vanishing extrinsic curvature\footnote{As we will see, surfaces with $K=0$ generally exist for real solutions of JT gravity. This is in sharp contrast to higher dimensional gravity theories in which the lack of such a $K=0$ surface may considered a primary obstruction to obtaining real wormhole solutions.  For JT, the obstruction is instead the desire to satisfy boundary conditions with the same sign of $\phi_b$ on each boundary.  In addition, for JT gravity the extrinsic curvature Poisson-commutes with the induced metric, whose conjugate is instead the normal derivative of the dilaton.}.  For the isolated $K=0$ throat considered below, the constraint keeps the metric, dilaton, and extrinsic curvature continuous while allowing a jump in the normal derivative of the dilaton.
To be definite, we consider the extrinsic curvature $K$ defined by the normal pointing in the direction in which the average value of $\chi$ increases.   We implement the desired Neumann constraint by adding a Lagrange multiplier term to the total action:
\begin{equation}
    S_{con} = \lambda_A \left[ \frac{1}{2}\int_{\mathcal{M}} d^2x \sqrt{g} n_\chi^\mu \nabla_\mu \text{sgn}(K_\chi) - A_0 \right],
    \label{eq:SconJT}
\end{equation}
yielding the complete action $S = S_{JT} + S_{matter} + S_{con}$.  This constraint contributes a new localized source term to the equation of motion for the dilaton field $\Phi$, where the source term is proportional to $\lambda_A$.  As a result, it allows a discontinuity in the derivative of the dilaton but requires continuity of the metric, dilaton, and extrinsic curvature.
We construct the relevant classical solutions in section \ref{sec:JTcl} before computing their action in section \ref{sec:JTS} and studying the resulting partition functions in \ref{sec:JTZ}.

\subsection{Constrained Classical solutions}
\label{sec:JTcl}

Extremizing the action with the constraint term yields the classical equations of motion for a constrained wormhole. In JT gravity, the dilaton equation of motion enforces a constant negative curvature geometry. This sets $R=-2$ everywhere. We adopt the Euclidean metric ansatz
\begin{equation}
    ds^2 = dr^2 + f(r)dt_E^2,
\end{equation}
where both the metric and matter fields are assumed to be invariant under translations of $t_E$ and where $t_E$ has real period $\beta$.
Following \cite{Held:2026bbo}, one can parameterize the general solutions to the resulting equations of motion equations using four integration constants ($c_1, c_2, c_3$, and $C_0$). The metric function takes the form
\begin{equation}
    f(r) = c_1^2(e^r + c_2^2e^{-r})^2.
\end{equation}
The scalar field and the dilaton profiles are given by
\begin{align}
    \chi(r) &= \frac{q}{c_1 c_2}(R(r) - R_0), \\
    \Phi(r) &= -\frac{q^2}{8c_1^2 c_2^3}\left[c_2 + (e^r - c_2^2 e^{-r})R(r)\right] - c_3(e^r - c_2^2 e^{-r}),
\end{align}
where we define the function $R(r) \equiv \arctan(e^r/c_2)$. The constant $R_0 \equiv \arctan(e^{r_0}/c_2)$ absorbs the global shift symmetry of the scalar field. This allows us to fix $\chi(r_0) = 0$ at some reference radius $r_0$.

The integration constants are determined by the boundary conditions and the geometric properties of the wormhole. First, we consider the scalar equation of motion:
\begin{equation}
    \chi''(r) + \frac{f'(r)}{2f(r)}\chi'(r) = 0.
\end{equation}
This equation implies the existence of a conserved quantity
\begin{equation}
 q:=C_0=\sqrt{f(r)}\chi'(r).
\end{equation}
Match this quantity with the Neumann boundary condition at asymptotic infinity yields $q=C_0$.

To understand the effect of introducing the constraint term $S_{con}$, we first identify the geometric throat of the wormhole. The throat corresponds to the minimum of the metric function $f(r)$; i.e., to the location where $f'(r_0) = 0$. This condition yields $e^{r_0} = c_2$. To make the $\mathbb{Z}_2$ symmetry of the wormhole solution explicit, we shift our radial coordinate to place the throat at the origin ($r_0 = 0$). This choice immediately sets $c_2 = 1$.

Recall now that the constrained action fixes the proper area of this throat to $A_0$. We evaluate the induced metric at the throat and integrate over $t_E$ to find:
\begin{equation}
   A_0 =  A(r_0) = \beta \sqrt{f(r_0)} = 2c_1 c_2 \beta.
\end{equation}
Since $c_2 = 1$, we find $c_1 = A_0 / (2\beta)$. Recall that, at this point, $A_0$ has simply been chosen arbitrarily.  We will later find that extremizing over $A_0$ will relate $A_0$ to $k$.

Finally, the remaining parameter $c_3$ is fixed by the  boundary conditions at infinity. Let us therefore consider the limit $r \to \infty$ in which the dilaton profile takes the asymptotic form:
\begin{equation}
    \Phi(r) \approx -\left(\frac{\pi q^2}{16 c_1^2} + c_3\right)e^r.
\end{equation}
We wish to match this asymptotic behavior to the regularized boundary value $\Phi_{bdy} = \phi_b/\epsilon$ at the cutoff surface. This matching completely determines $c_3$. Since $\sqrt h\sim c_1e^r=\epsilon^{-1}$, the dilaton boundary condition requires
\begin{equation}
    -\left(\frac{\pi q^2}{16c_1^2}+c_3\right)=\phi_b c_1.
\end{equation}
For $\mathbb Z_2$-symmetric wormholes we find
\begin{equation}
    k=\chi_R-\chi_L=\frac{\pi q}{2c_1},
    \qquad
    q=\frac{A_0k}{\pi\beta},
    \label{eq:JTqk}
\end{equation}
so that the integration constants take the values
\begin{equation}
    c_1=\frac{A_0}{2\beta},\qquad
    c_2=1,\qquad
    c_3=-\frac{\phi_bA_0}{2\beta}-\frac{k^2}{4\pi}.
    \label{eq:JT-constrained-family}
\end{equation}
Thus $A_0$ remains an independent constrained variable. 

For completeness, the dilaton derivative at the gluing surface (evaluated on the right-hand side) is
\begin{equation}
    \Phi'_+(0)=2\phi_b c_1+\frac{k^2}{4\pi}
    =\frac{\phi_bA_0}{\beta}+\frac{k^2}{4\pi},
\end{equation}
while evaluating on the left-hand side gives $\Phi'_-(0) = - \Phi'_+(0)$.
We remind the reader that this jump in $\Phi'$ must be proportional to $\lambda_A.$ Requiring $\lambda_A$  to vanish, or equivalently requiring a smooth unconstrained dilaton, gives
\begin{equation}
    A_0=-\frac{\beta k^2}{4\pi\phi_b},\qquad
    c_1=-\frac{k^2}{8\pi\phi_b},\qquad
    c_3=-\frac{k^2}{8\pi}.
    \label{eq:JT-smooth-constants}
\end{equation}
In particular, since the parameter $A_0$ appears in the action only through \eqref{eq:SconJT}, and since extremizing \eqref{eq:SconJT} with respect to $A_0$ clearly sets $\lambda_A=0$, 
the above relations must also hold if we extremize over $A_0$.  In particular, setting $k=2ik_E$ we then recover the smooth Euclidean wormhole of \cite{Garcia-Garcia:2020ttf} (without the one-loop factors studied there): $c_1=k_E^2/(2\pi\phi_b)$ and $S=-\beta k_E^4/(\pi^2\phi_b)$.

\subsection{The action}
\label{sec:JTS}

We now evaluate the on-shell action $S_{CWH} = S_{JT} + S_{matter}$.  First, we consider the gravitational sector. Since the dilaton equation of motion enforces the constant negative curvature condition $R=-2$, the bulk integral in the JT action vanishes entirely. The gravitational action is therefore determined only by the boundary terms.  These two terms agree due to the $\mathbb{Z}_2$ symmetry and yield:
\begin{equation}
    S_{JT} = -2 \int_0^\beta dt_E \left[ \sqrt{h}\, \Phi (K-1) \right]_{r \to \infty}.
\end{equation}
To compute the result, we use the asymptotic expansion of the extrinsic curvature:
\begin{equation}
    K-1 \approx {-2e^{-2r}}.
\end{equation}
We also use the dilaton profile matched to the cutoff value $\Phi_{bdy} = \phi_b/\epsilon$.  These expressions ensure the expected cancellation of divergences, leaving a finite result that can be written in terms of the throat area $A_0$ and the thermal period $\beta$: \begin{equation}
    S_{JT} = \frac{\phi_b}{\beta} A_0^2.
\end{equation}

Next, we turn to the matter sector. The relevant action is the sum of the bulk scalar action and the Neumann boundary term. Using the $\mathbb{Z}_2$ symmetry of the wormhole and flux conservation we find the bulk term in \eqref{eq:SmatterJT} to be
\begin{equation}
    S_{matter, \ bulk} = 2 \times \frac{1}{2} \int_0^\beta dt_E \int_0^\infty dr \sqrt{f(r)}(\chi'(r))^2
    =\beta q\chi(\infty)=\frac{\beta qk}{2}.
\end{equation}
Since $k$ and $q$ have the same sign by \eqref{eq:JTqk}, we see that the above Dirichlet  matter action is positive for real $k$.

As noted earlier, a well-defined Neumann variational problem requires the inclusion of the boundary term $S_{bdy, \chi}$, which yields
\begin{equation}
    S_{bdy, \chi} = -\int_{\partial \mathcal{M}} dt_E \sqrt{h}\, \chi \, n^\mu \partial_\mu \chi
    =-2\beta q\chi(\infty)=-\beta qk,
\end{equation}
where the factor of $2$ comes from the fact that $\partial {\cal M}$ includes both the left and right boundaries.
 The Neumann matter action is thus given by:
\begin{equation}
    S_{matter}^{(N)} = S_{matter, \ bulk} + S_{bdy, \chi} = -\frac{\beta qk}{2}  = -\frac{\pi\beta^2 q^2}{2A_0},
\end{equation}
where the last step used $q=C_0$ and $k=\pi\beta q/A_0$.

Finally, let us recall that we have enforced the constraint obtained by varying $S_{con}$ with respect to $\lambda_A$.    As a result,  $S_{con}$ now vanishes.  Adding the gravitational and matter terms thus gives the total Euclidean on-shell action.  For Neumann boundary conditions we find :
\begin{equation}
\label{eq:CWHN}
    S_{CWH}^{(N)}(A_0, \beta;q) = \frac{\phi_b}{\beta} A_0^2-\frac{\pi\beta^2 q^2}{2A_0}.
\end{equation}

For comparison, using $\chi_R=-\chi_L = k/2$, with Dirichlet boundary conditions we obtain 
\begin{equation}
\label{eq:JTSD}
    S_{CWH}^{(D)}(A_0,\beta) = \frac{\phi_b}{\beta}A_0^2 + \frac{k^2}{2\pi}A_0.
\end{equation}
in agreement with \cite{Held:2026bbo}.
The fact that matter term is proportional to $A_0$ in the Dirichlet action and to $A_0^{-1}$ in the Neumann action will lead to rather different structures when we integrate over $A_0$.

\subsection{The two-boundary partition function}
\label{sec:JTZ}

As noted above, our general strategy for studying the two-boundary partition function follows that of  \cite{AxionWormholes,Held:2026bbo}.  In particular, we assume that the full partition function can be written as an integral over $A_0$ of partition functions defined at fixed $A_0$.  We then assume that we may approximate such fixed-$A_0$ partition functions by $e^{-S_E}$ evaluated on the constrained saddles studied above.  After doing so, we may then study the resulting partition function by again using the saddle-point approximation to perform the remaining integral over $A_0$, though we will include a full Picard-Lefschetz analysis of that integral to ensure that we include the appropriate set of saddles. 
In particular, we assume this final partition function to be defined by integrating $A_0$ over the domain defined by real $A_0 \ge 0$.

The Dirichlet case, 
\begin{equation}
\label{eq:ZD}
Z_D(k,\beta) : = \int_{A_0>0} dA_0 \ e^{-S_{CWH}^{(D)}},
    \end{equation}
was previously analyzed in \cite{Held:2026bbo}, but let us take a moment to briefly recall the results. 
Let us first consider the case of real $k$. The action  (\eqref{eq:JTSD}) has no stationary point at positive $A_0$.  Instead, a stationary point would require 
\begin{equation}
    A_0=-\frac{\beta k^2}{4\pi\phi_b},
\end{equation}
which has $A_0< 0$. Since $A_0$ sets $\lambda_A=0$,  the resulting  formal solution satisfies the smoothness condition (\ref{eq:JT-smooth-constants}), though of course the solution is not in fact smooth since reaching negative $A_0$ requires $f$ to pass through zero.  

Since the saddle has $A_0 <0$, it does not lie on the contour that defines our integral.  Indeed, it lies on the opposite side of the endpoint $A_0=0$.  One may thus expect that it does not contribute.  That this is so can be seen explicitly from the fact that for $A_0>0$ \eqref{eq:JTSD} is a strictly increasing function of $A_0$, so that the integral over $A_0 >0$ is dominated by the endpoint at $A_0=0$.  However, as noted in \cite{Held:2024qcl}, one may then consider analytically continuing $k$ into the complex plane.  Upon doing so, one finds that the integral over positive real $A_0$ remains convergent and gives the desired analytic continuation.  As a result, after continuing $k$ to imaginary values so that the saddle moves to the positive real axis, one finds that it now lies on the integration contour and that it is a local minimum of the action along this contour. It thus clearly contributes to the partition function.

Indeed, as discussed in \cite{Held:2026bbo}, the saddle in fact contributes for $|{\rm Im} k| > |{\rm Re} k|$.  Furthermore, it dominates over the endpoint contribution when $k = \pm i e^{i\alpha} |k|$ with  $|\alpha| < \frac{\pi}{8}.$

Let us now turn to the Neumann case.  In a naive Euclidean approach one might expect to define the Neumann partition function $Z_N(q)$ by 
taking the flux $q$ to be real and 
integrating $e^{-S_{CWH}^{(N)}}$ over positive real values of $A_0$.  However, one sees from \eqref{eq:CWHN} that $S_{CWH}^{(N)} \rightarrow -\infty$ as $A_0 \rightarrow 0$ along the positive real axis.  As a result, such an integral would diverge exponentially.

As foreshadowed in the introduction, this exponential divergence in fact stems from the fact that the energy of our constrained wormholes is unbounded below.  To see this connection we need only note that, due to the nowhere-vanishing hypersurface-orthogonal Killing field associated with Euclidean time translations, on general grounds (see e.g. \cite{Carlip:1993sa}) the Euclidean action is just $S_{CWH}^{(N)} = \beta E$ where $E$ is the energy conjugate to the corresponding Lorentz-signature time-translations.  Since $\beta$ is held constant, we then find $E = S_{CWH}^{(N)}/
\beta \rightarrow -\infty$ as $A_0\rightarrow 0$.

On the other hand, the corresponding integral converges when $|{\rm Im} \ q| > |{\rm Re} \ q|.$  This is consistent with our analysis of the Dirichlet partition function $Z_D$ which at large real $k$ was given by an endpoint contribution that decays only like a power law in $k$.  As a result, while the Fourier transform of $Z_D$ is well-defined, the integral of $Z_D$ against $e^{\beta qk}$ diverges as  $k \sign (q) \rightarrow + \infty$.   As stated in the introduction, we will therefore simply write $q=iQ$, define
\begin{equation}
Z_N(q=iQ) : = \int_{A_0 > 0}dA_0 \ e^{-S_{CWH}^{(N)}(A_0,\beta;q=iQ)}
\label{eq:JTNint}
\end{equation}
for real $Q$, and then analytically continue the result to imaginary values $Q = -i q$ (so that $q$ is then real).  

\begin{figure}[tbp]
    \centering
    \begin{minipage}[t]{0.48\linewidth}
        \centering
        \includegraphics[width=\linewidth]{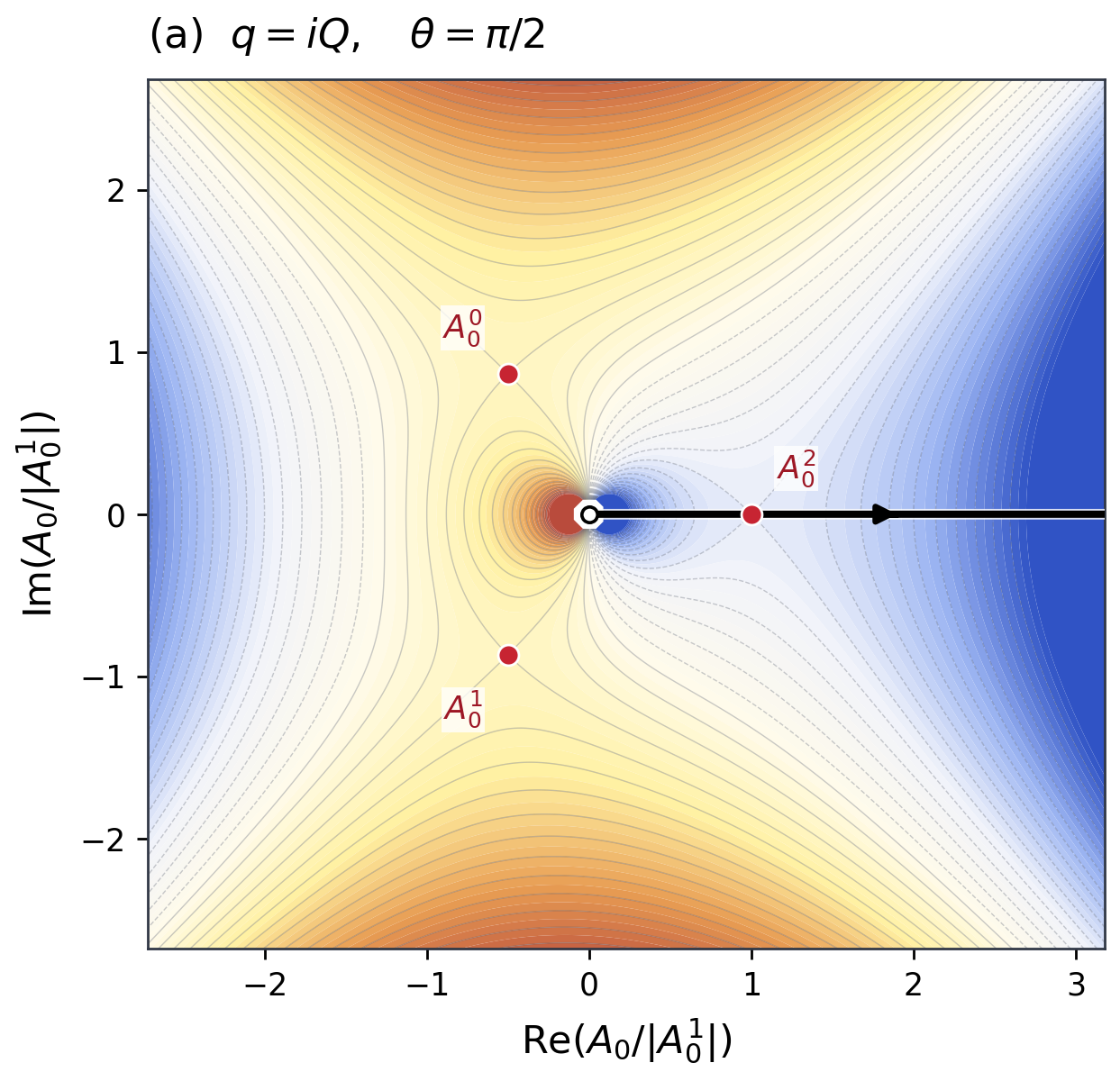}
    \end{minipage}\hfill
    \begin{minipage}[t]{0.48\linewidth}
        \centering
        \includegraphics[width=\linewidth]{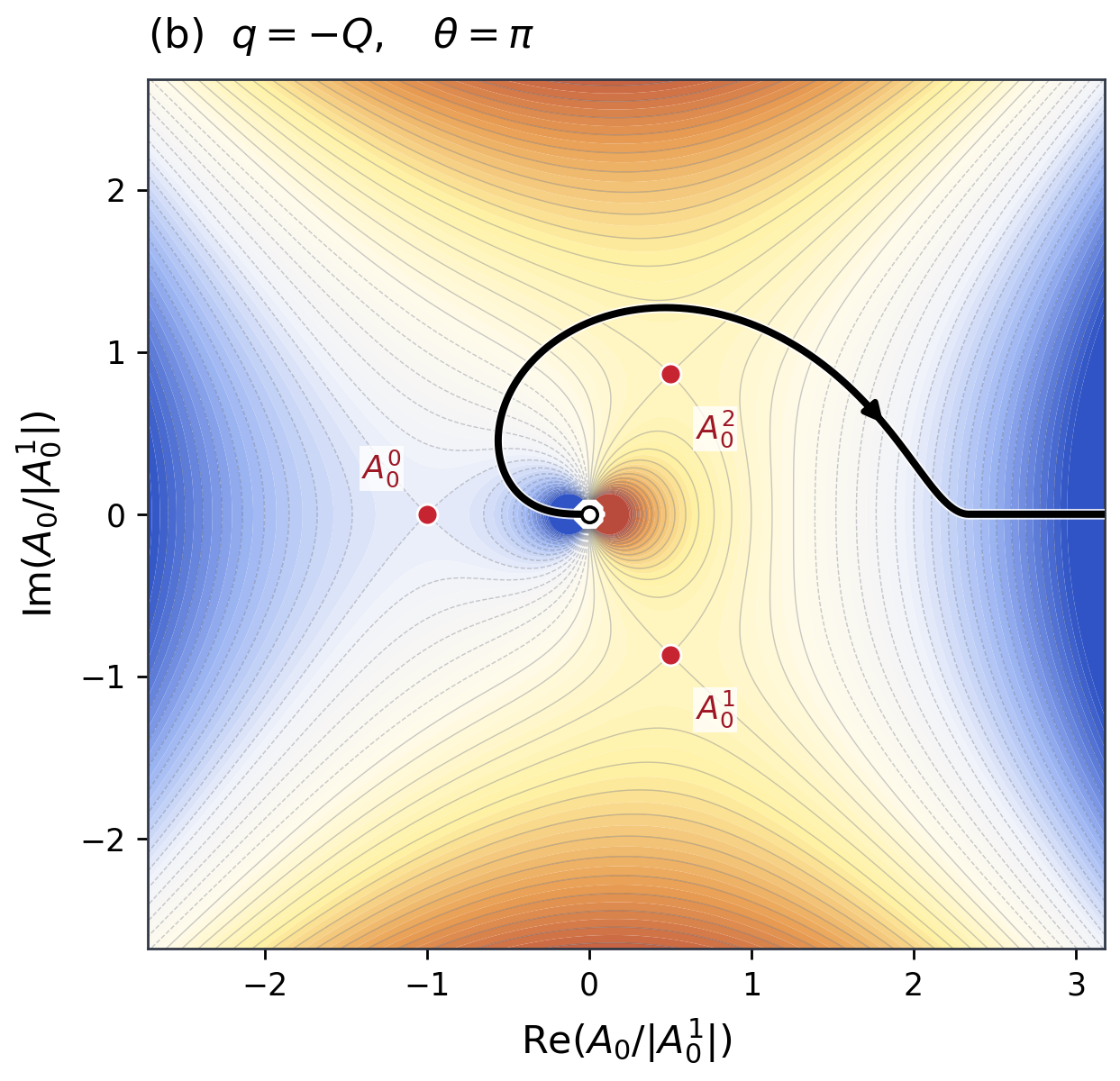}
    \end{minipage}

    \par\vspace{0.6em}

    \begin{minipage}[t]{0.48\linewidth}
        \centering
        \includegraphics[width=\linewidth]{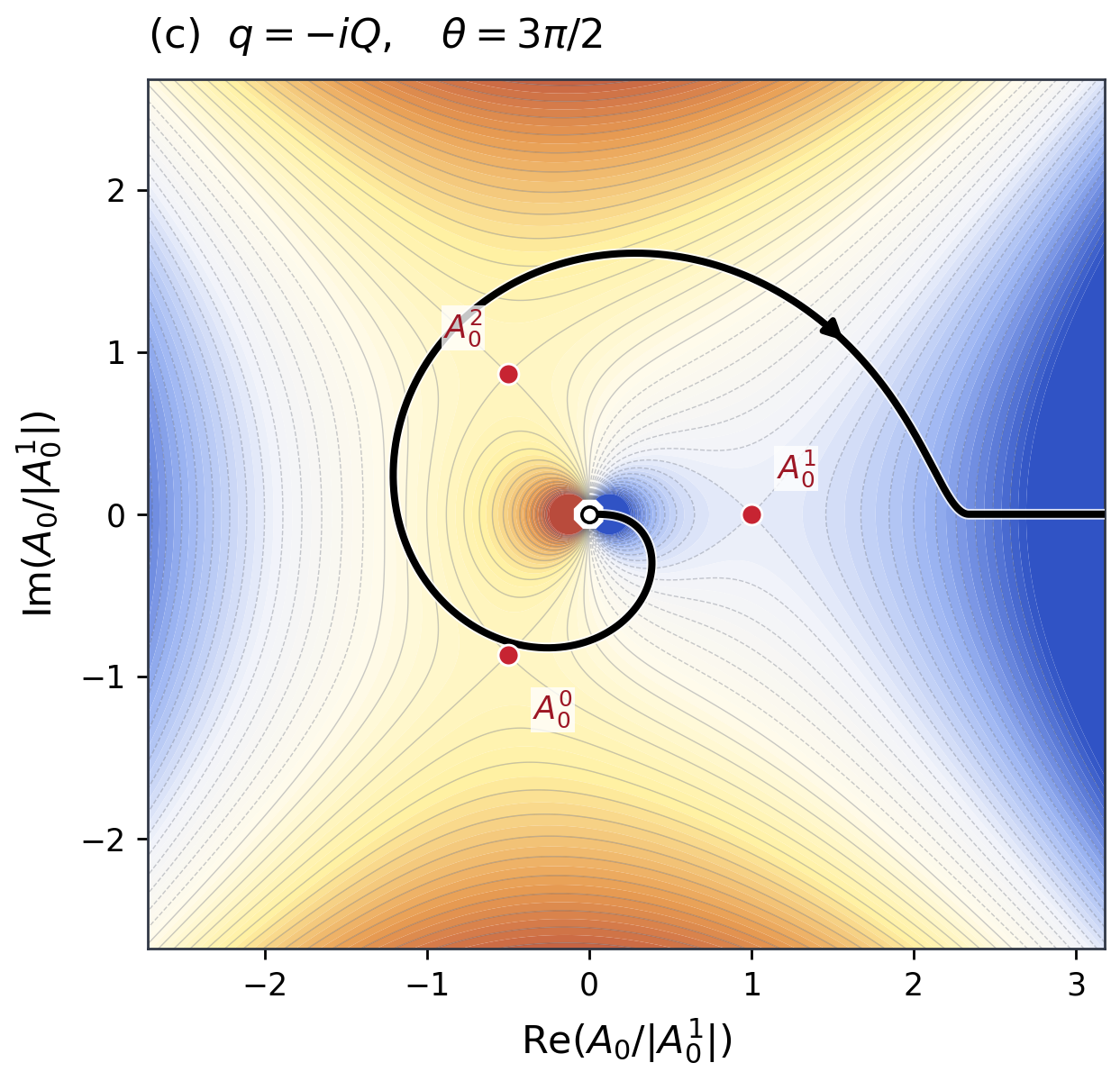}
    \end{minipage}\hfill
    \begin{minipage}[t]{0.48\linewidth}
        \centering
        \includegraphics[width=\linewidth]{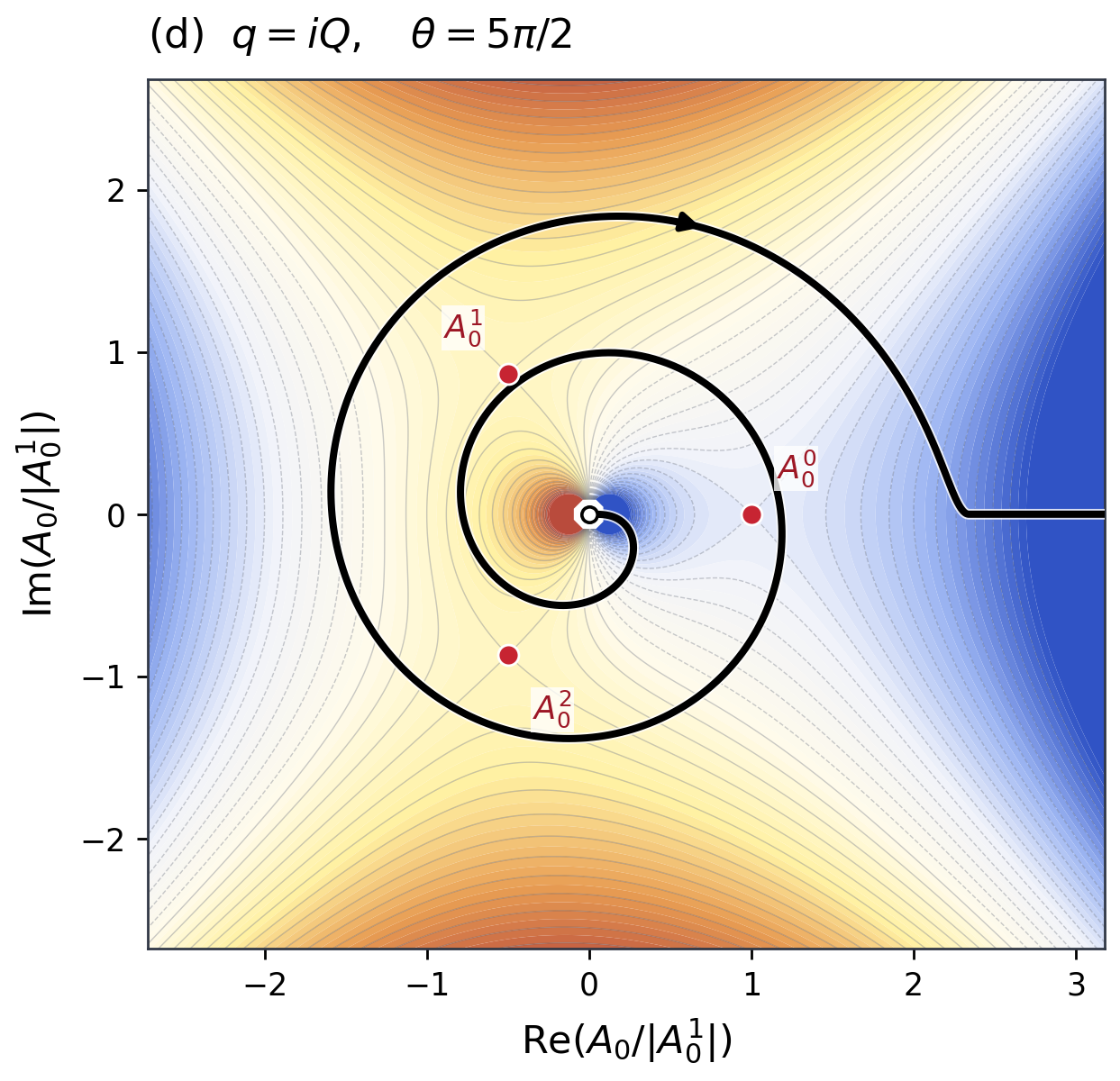}
    \end{minipage}

    \caption{Integration contours in the complex $A_0$ plane at the indicated
values of the continuous phase $\theta=\arg q$, analytically continued from
$q=iQ$ with $Q>0$. Black arrows indicate contour orientation, red dots mark
the saddles $A_0^n$, and the shading shows $\operatorname{Re}(-S_{CWH}^{(N)})$ with larger values in warmer tones. The axes are in units of $|A_0^1|$.}
    \label{fig:windingcontours}
\end{figure}

This final step can be enacted by deforming the $A_0$ integration contour so that the direction from which it approaches $A_0=0$ varies with the phase of $q$.  In particular, if we define $Z_N(q)$ by analytic continuation of \eqref{eq:JTNint} starting at values of $q$ on the positive imaginary axis, we may choose the contour near $A_0=0$ to have 
\begin{equation}
    \label{eq:argA0}
{\rm arg}(A_0) = 2\, {\rm arg}(q) - \pi.
\end{equation}
Monotonically increasing the phase of $q$ thus has the effect of winding our contour around the essential singularity at $A_0=0$; see figure \ref{fig:windingcontours}. 

\begin{figure}
                \centering
\includegraphics[width=0.6\linewidth]{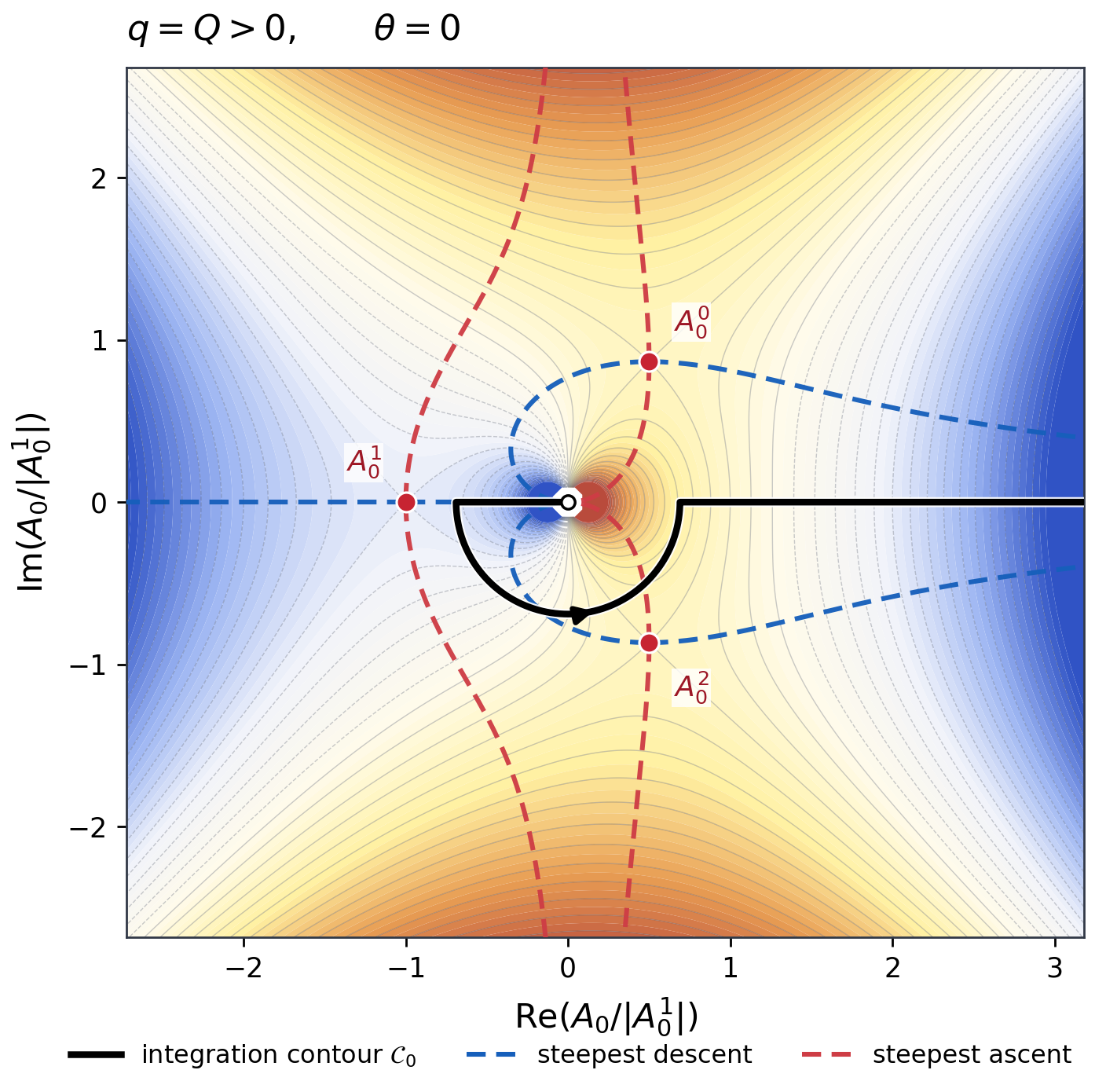}
                \caption{The $q>0$ with $\theta=0$ contour obtained by clockwise continuation
from $q=iQ$. The black curve is the integration contour; blue and red dashed
curves are steepest-descent and steepest-ascent contours, respectively. Red
dots mark the saddles $A_0^n$, and the shading shows
$\operatorname{Re}(-S_{CWH}^{(N)})$. The axes are in units of $|A_0^1|$.}
                \label{fig:JTqreal}
                \end{figure}

Let us now consider the stationary phase approximation to the resulting $Z_N$ for real positive $q$.  Figure \ref{fig:JTqreal} shows the three saddle points of $S_{CWH}^{(N)}$ at
\begin{equation}
A_0 = \left( -\frac{\pi q^2 \beta^3}{4 \phi_b}\right)^{1/3}  \ \ \ {\rm with}  \ \ S_{CWH}^{(N)} = -\frac{3\pi\beta^2 q^2}{4A_0}.
\label{eq:JTNsaddles}
\end{equation}
along with the associated contours of steepest descent and ascent and the contour of integration.  The shading indicates the magnitude of $S_{CWH}^{(N)}$ in various regions of the $A_0$-plane.  We see that only the saddle $A_0^2$ (lower right) has non-zero intersection number with our contour.  Indeed, the integration contour can be smoothly deformed to the steepest descent contour through $A_0^2$.  Since the $S_{CWH}^{(N)} \rightarrow +\infty$ as one approaches $A_0=0$ along the integration contour there can be no endpoint contribution.

\subsection{General complex $q$}

As noted above, we may also consider analytic continuation to other values of $q$.  As we do so, we encounter Stokes' lines where the ascent flow from one saddle intersects another saddle. The presence of such transitions can be seen from figure \ref{fig:windingcontours}.  These turn out to occur every time two saddles have the same value of ${\rm Im} \ S_{CWH}^{(N)}$, which from \eqref{eq:JTNsaddles} we see occurs when ${\rm Im} (q^{4/3} \omega^n)$ has the same value for two values of $n$ where $\omega = \frac{\sqrt{3} i - 1}{2}$.  A short calculation then shows that this occurs when ${\rm arg}\  q^{4/3}$ is an odd multiple of $\frac{\pi}{6}$, and thus when ${\rm arg}\  q$ is an odd multiple of $\frac{\pi}{8}$.  

Since the action $S_{CWH}^{(N)}$ depends on $q$ only through $q^2$, the entire pattern of ascent/descent contours is manifestly periodic after a rotation of the $q$-plane by $\pi$.  There are thus four geometrically-distinct cases to be considered. Since we have no strong interest in the answer for any particular complex value of $q$, we will not discuss these cases in detail.  However, it is clear that the descent thimbles through each saddle will sometimes run between the asymptotic regions at large real $\pm A_0$ and will sometimes run from one of those regions to the origin.  For the interested reader, we simply provide a table of such thimbles in \eqref{eq:JT-Stokes-endpoints} at values of ${\rm arg} \ q$ representing each of the cases.  Here we have labeled the saddles $i=0,1,2$ by writing $q = Qe^{i \theta}$ for real $Q, \theta$ and taking the saddle points to be 
$A_0^n = e^{i\frac{2\theta+(2n+1)\pi}{3}} \left(\frac{\pi Q^2 \beta^3}{4 \phi_b}\right)^{1/3}.$  The table lists the initial and final endpoints of each thimble (defined according to the orientation), using the symbol $0$ for the essential singularity at the origin and $\pm$ for the asymptotic regions at positive or negative real $A_0$.
\begin{equation}
\begin{array}{c|c|c|c}
 \frac{8{\rm arg} \ q}{2\pi}  & J_0 & J_1 & J_2 \\ \hline
 0 & 0\to+ & -\to0 & 0\to+ \\
 1 & -\to+ & -\to0 & 0\to+ \\
 2 & -\to0 & -\to0 & 0\to+ \\
 3 & -\to0 & -\to+ & 0\to+ \\
 4 & -\to0 & 0\to+ & 0\to+ \\
\end{array}
 \label{eq:JT-Stokes-endpoints}
\end{equation}

To understand how these transitions interact with the winding of the contour shown in figure \ref{fig:windingcontours}, let us consider what happens when $q$ circles the origin once (say, in the counter-clockwise direction).  From (\ref{eq:JTNsaddles}) we find  $A_0^n(\theta+\pi)=A_0^{n+1}(\theta)$ with indices defined modulo three. As a result, after half a circuit of the $q$-plane, the saddle points and their descent contours return to the same geometric positions, though the labels assigned to those positions have been permuted cyclicly. In particular, the rows labeled $0$ and $4$ in table \eqref{eq:JT-Stokes-endpoints} differ only by such a cyclic permutation.

From \eqref{eq:argA0}, we see that for every half-circuit of the $q$-plane, the direction of the contour near $A_0=0$ changes by $2\pi$, rotating once around the origin in the counter-clockwise direction, though the large-$A_0$ end of the integration contour remains on the positive real axis. As a result, after $m$ full circuits around the origin of the $q$-plane, for any saddle whose ascent contour reaches the origin the intersection number with the integration cycle has changed by $2m$.  (In contrast, for ascent contours that stretch between the two asymptotic regions, the net signed intersection number remains zero.) 

The above linear growth with $m$ indicates the possibility that $Z_N$ may have a logarithmic singularity at $q=0$.  However, whether such a singularity in fact arises depends on the choice of measure in \eqref{eq:JTNint}, which is not determined by our leading-order semiclassical analysis.  In particular, had we chosen to define the Dirichlet partition function $Z_D$ using the measure in \eqref{eq:ZD}, and to then define $Z_N$ as the analytic continuation of its Fourier transform, the transform of the Gaussian factor in \eqref{eq:ZD} would have introduced an extra factor of $A_0^{-1/2}$ in the measure of \eqref{eq:JTNint}.  The integrand would then be properly defined on a Riemann surface that covers the $A_0$-plane twice.  On that Riemann surface, we then find {\it six} saddles that come in pairs with opposite signs of $A_0^{-1/2}$.  
Since the action depends on $q$ only through $q^2$, the pattern of such saddles is again periodic when ${\rm arg} \ q$ changes by $\pi$. 
However, it now requires a full circuit around the origin of the $q$-plane (with $\arg (q)$ changing by $2\pi$) for the direction of the contour at $A_0=0$ to fully wind around the origin of the above Riemann surface.

After each such period, the resulting winding of our contour thus changes the intersection number by one for each of the six saddles whose ascent contour runs to the origin.  The key point is then that the corresponding contributions to $Z_N$ then cancel in pairs associated with opposite signs for $A_0^{-1/2}$.  So in this case the final $Z_N$ has no logarithmic singularity, and in fact no branch points in the $q$-plane.  

We emphasize again that our leading-order semiclassical analysis simply does not determine such factors in the measure, so that the above discussion should be taken as merely an illustrative example of what might occur if such factors were included. It would thus be interesting to return to this case and include appropriate one-loop corrections in future work.

On the other hand, noting that there is indeed the possibility of a logarithmic singularity in $Z_N$, let us take a moment to understand how that can be compatible with the manifest lack of any such singularity in $Z_D$ and, in particular, with the fact that $Z_D$ undergoes only a simple Stokes' transition that turns the contribution of the saddle on and off.   To do so, let us imagine computing $Z_D$ as above (including, perhaps, an arbitrary measure factor) and then computing its Fourier transform $\tilde Z_D(\mathcal Q)$, with $Z_{N}(q)$ at real positive $q$ given by evaluating $\tilde Z_D(\mathcal Q)$ at $\mathcal Q=-i\beta q$ (so that $\beta q = i{\mathcal Q}$).   Since the Fourier transform involves $e^{ik\mathcal Q}$, analytic continuation away from real $\mathcal Q$ generally requires that we deform the contour over which we integrate $k$.  But when $k$ approaches the imaginary axis, we saw that the saddle contributes to $Z_D$, and that it is exponentially large in $k^4$ when ${\rm arg} \ k$ is within $\pi/8$ of this axis.  As a result, once this exponential growth becomes relevant, it will simply not be possible to maintain convergence by further deforming the contour over which we integrate $k$.  Notably, this occurs even before we can take $q=iQ$ to be real, so this strategy simply fails to define the analytic continuation of interest, even though it can be defined by performing the $k$-integral before performing the $A_0$-integral to write $\tilde Z_D$ in the form \eqref{eq:JTNint} discussed above for $Z_N$ and by choosing an appropriate $A_0$ contour as above.

This then resolves the apparent tension with the simple character of Stokes' transitions for $Z_D$.  It also shows that, even this simple JT gravity wormhole gives an example where the contributions of saddles to $Z_N$ cannot be deduced in a simple way from the contributions of what may appear to be the same saddles of $Z_D$.  In particular, using the saddle-point relation $k^3=  -4\pi^2 q \phi_b$ to map between $k$ and $q$, after winding $q$ once around the origin the wormhole can clearly contribute even in a regime where no wormholes contribute to $Z_D$.

\subsection{Other examples of Dirichlet/Neumann wormhole mismatch}

While the potential logarithmic winding may be the most dramatic example of smooth wormhole saddles contributing in different ways to $Z_D$ and $Z_N$, one can also find examples
that arise when there are no branch points in the $q$-plane; e.g., when the extra factor of $A_0^{-1/2}$ mentioned above is included in the integration measure for $Z_N$.

We will mention two such example here (though this list is not exhaustive).  The first occurs for real $q$ (i.e., for $\theta=0$).  As can be seen from figure \ref{fig:JTqreal}, the ascent curve for the upper right saddle ($A_0^0$) has zero intersection number with the contour of integration for $Z_N$.  So it does not contribute to $Z_N$.  Yet from \eqref{eq:JTqk}  this saddle has $\arg(k) = -\pi/3$, and thus $|\arg (k)|> \pi/4$, so it contributes to $Z_D$.

Another example occurs at negative imaginary $q$ (i.e., for $\theta = 3\pi/2$).  Examining the bottom left panel in figure \ref{fig:windingcontours}, one sees that the ascent curves for the two left saddles both flow to the origin, and thus have intersection number $\pm 1$ with the contour of integration for $Z_N$.  So these saddles contribute to $Z_N$.  While each contribution comes with some yet-to-be determined sign from the factor of $A_0^{-1/2}$, there is only one such contribution for each saddle shown, so we cannot find pairs of canceling saddles associated with the two branches of $A_0^{-1/2}$.  Furthermore, the two saddles that do contribute have different semiclassical actions and so cannot cancel against each other.  But \eqref{eq:JTqk} then gives $\arg (k)=\pi/6, 5\pi/6$ so $|{\rm Re}\, k|> |{\rm Im}\, k|$ and these wormholes do not contribute to $Z_D$.

\section{alAdS Axion Wormholes with Toroidal Boundaries}
\label{sec:tori}

We now perform a similar analysis in an asymptotically locally AdS setting.  To be definite, we work in $d=3$ spacetime dimensions.  In order to continue to focus on cases where gravitational entropy is unimportant (and where the Euclidean action is simply $\beta E$), we take the conformal boundary to be a torus.  As in section \ref{G3wormhole}, we also include a minimally-coupled scalar $\chi$.  In the minimally-coupled case, this $\chi$ may be equally-well called an axion, so that we consider toroidal axion wormholes.   This contrasts with the case of spherical asymptotically-AdS$_3$ axion wormholes studied in \cite{AxionWormholes}.

We will again impose the constraint that the  area of the (codimension-1) $\chi=0$ locus takes some fixed value $A_0$. As section \ref{G3wormhole}, imposing this constraint adds a source to the equations of motion that is equivalent to introducing a codimension-1 membrane on the surface where the constraint is imposed. The contribution of the membrane to the Euclidean action $S_E$ is the same as what would be obtained by cutting the geometry at the resulting two new finite boundaries and adding boundary terms there as required to fix $\chi=0$ and $A=A_0$.  We thus find\footnote{Our normalization of $\chi$ here and in section \ref{sec:asFlat} differs from that of section \ref{G3wormhole} by a factor of $\sqrt{2} \kappa$.} 

        \begin{equation}
          S_E=  -\frac{1}{2\kappa^2}\int_\mathcal{M}d^3x\,\sqrt{g}\,\Big(R+\frac{2}{\ell^2} - \frac{1}{2}\partial_\mu\chi\partial^\mu\chi\Big) - \frac{1}{\kappa^2}\int_{\partial\mathcal{M}}d^2x \,\sqrt{h}\,K \, +S_{\text{ct}} +S_{\chi_\infty},
        \end{equation}
            where \begin{equation}S_\text{ct} = \frac{1}{\kappa^2\ell}\int_{\partial\mathcal{M}}d^2x\,\sqrt{h}\end{equation} is the usual counterterm for alAdS$_3$ spacetimes (since the would-be curvature counterterm vanishes for boundary tori) and $S_{\chi_\infty}$ is the boundary action required to obtain a good variational principle under variations of the axion. For Dirichlet and Neumann boundary conditions, this is
        \begin{equation}
        \label{eq:SbndyDN}
            S_{\chi_\infty}=
            \begin{cases}
                 0, &\text{(Dirichlet)}\\
                -\frac{1}{2\kappa^2}\int_{\partial\mathcal{M}}d^2x\,\sqrt{h} \,\,\chi n^\mu\partial_\mu\chi, &(\text{Neumann)}
            \end{cases}
        \end{equation}
    where $n^\mu$ is the outward directed unit normal to the boundary.  

    Let us take the following ansatz for the metric for our constrained toroidal wormholes:

    \begin{equation}
        ds^2 = dr^2 +\frac{A(r)}{2\pi\beta}\big( d\theta^2 +d\tau^2\big).
    \end{equation}
    Here $r\in(-\infty,\infty)$ and $\theta$ is $2\pi$ periodic, while $\tau$ has a period of $\beta>0$.  We can additionally find the axion flux density by using the conserved current associated with uniform shifts of the axion, 
    \begin{equation}
    \begin{split}
 q&=\frac{1}{2\pi\beta}\int_{r=\mathrm{const.}}d^2x\sqrt h \,n^\mu\partial_\mu\chi\\
 &=\frac{1}{2\pi\beta}A(r)n^\mu\partial_\mu\chi(r)\quad\Longrightarrow\quad
 \partial_\mu\chi=\frac {2\pi\beta q}{A(r)}\delta_\mu^r.
\end{split}
\end{equation}
    We fix the boundary conditions for the metric by expanding in some appropriate Fefferman-Graham coordinate at large $|r|$ that we call $z$. At a cutoff surface $z=\delta$, we impose

    \begin{equation}h_{ij}\,dx^idx^j\Big|_{z=\delta} = \frac{\ell^2}{\delta^2}(d\theta^2 +d\tau^2) +\mathcal{O}(\delta^0),
    \end{equation}
    for $\delta\ll1$.  
    We can now solve the rr Einstein equation to obtain an expression for A(r), 
    \begin{equation}
    \begin{aligned}
 A'^2&=4A\left(\frac{\pi^2 \beta^2 q^2}{A}+\frac{A}{\ell^2}\right)\\
 \Longrightarrow\quad A(r)&= \pi \ell  \beta q\sinh(2r/\ell)\qquad(q>0).
 \label{eq:AT1}
\end{aligned}
\end{equation}
Here we have imposed the above boundary condition for the metric at $r=+\infty$ by matching the expansion of the solutions to the equation of motion to the desired expansion of the metric.\footnote{In order to do this, we note that the appropriate Fefferman-Graham coordinate for $r>0$ is $z= z_0\,e^{-r/\ell}$. Expanding the solutions to the stated equation of motion for the metric and the matching boundary condition tells us that $z_0 =2\sqrt{\ell/q}$. This can be additionally used to identify the appropriate boundary cutoff in the proper length coordinate $r$. If the regulated boundary is taken to be at $z=\delta$, then that corresponds to $r=-\frac{\ell}{2}\log\!\left(\frac{q\delta^2}{4\ell}\right).$} 
However, the desired boundary condition at $r=-\infty$ is distinctly {\it not} satisfied since \eqref{eq:AT1} gives negative values of $A(r)$ for $r<0$.

In the final line of \eqref{eq:AT1} we have displayed the solution that is real and positive for $q>0$ and $r>0$.  However, since we will eventually be interested in complex solutions with $q$, it is useful to mention that with a different  (complex) constant of integration we may equally well take the solution of \eqref{eq:AT1} to be
    \begin{equation}
    A(r)=\pm i \pi\ell\beta q\cosh(2r/\ell).
 \label{eq:AT2}
\end{equation}

Recall, however, that we imposed a constraint on the codimension-1 of the surface $\chi=0$.  As noted above, this allows the presence of a source at that surface, so that constrained solutions can be constructed by cutting in half two of the above  one-boundary solutions and stitching together the two pieces on which $A>0$.  In particular, the constrained equations of motion require continuity of the metric, but allow the extrinsic curvature to be discontinuous at the stitching surface.

    As a reminder, the constraint we are imposing is that the area of the surface $\chi = 0$ takes some fixed value we will call $A_0$. Since surfaces of constant $\chi$ are surfaces of constant $r$ in our solutions $\exists \,r_0 \quad \text{s.t.} \quad \chi(r_0)=0$. Our constraint can then be written as $A(r_0)=A_0$. Using this form of the constraint and the solutions above, we can determine the axion profile,

    \begin{equation}
    \chi(r)=\int_{r_0}^r dr\,\partial_r\chi
 =\log\!\left[\coth(r_0/\ell)\tanh(r/\ell)\right].
\end{equation}
    We can then use this to relate the asymptotic value of $\chi$ to $r_0$,
    \begin{equation}
    \begin{aligned}
 \chi_\infty&=\chi\!\left(-\frac\ell2\log\!\left(
 \frac{q\delta^2}{4 \ell}\right)\right)\\
 &=\log\coth(r_0/\ell)+\mathcal O(\delta).
\end{aligned}
\label{eq:chiRelay}
\end{equation}
    Additionally, we can use our solution for $A(r)$ to relate $r_0$, $q$, and the value of the constrained area $A_0$.
    \begin{equation}
    A_0=A(r_0)=\pi\ell\beta q\sinh(2r_0/\ell).
 \label{eq:A0Relay}
\end{equation}
    Finally, with the solutions to all equations of motion in hand, we can proceed with finding the action for the constrained wormhole geometries of interest.

\subsection{The Constrained Wormhole Action}
    Let us first compute the action in terms of $\chi_R$, $q$, $r_0$, and $A_0$. Notably, $r_0$ will depend on the boundary conditions for the axion. For the Dirichlet boundary condition, it will depend on $\chi_R$ and $A_0$, as will $q$. However, with Neumann boundary conditions, both $r_0$ and $\chi_R$ are functions of $A_0$ and the boundary condition $q$. For simplicity we will restrict ourselves to solutions with $\chi_R = -\chi_L$, however this is certainly not necessary and one can obtain solutions with non-symmetric axion profiles by performing an overall shift of a symmetric one, though in those cases one must take care to enforce the constraint along the surface $\chi = \frac{\chi_R +\chi_L}{2}$ rather than at $\chi=0$ as we do here. For the problem with Neumann boundary conditions for the axion, this amounts to fixing the zero mode associated with overall shifts since the boundary values of the axion are not fixed.

    As a reminder, we will construct constrained wormholes by stitching together two real one-sided geometries along the codimension-1 constraint surface. The non-smoothness across the constraint surface induces a delta function in the Ricci scalar on the stitching surface. However, ignoring this contribution for now, the action of a one sided solution composing half of the constrained wormhole (WLOG say the ``right'' side) is,

    \begin{equation}
    S_R=-\frac{\pi\beta q}{\kappa^2}\cosh(2r_0/\ell)+S_{\chi_\infty,R}.
\end{equation}
    where $S_{\chi_\infty,R}$ is the boundary term \eqref{eq:SbndyDN} required to get a good variational principle at the right asymptotic boundary.  To obtain the first term, note that the trace of the Einstein equation gives $R=-6/\ell^2+\frac12(\partial\chi)^2$, so that the bulk Lagrangian takes the value $-4/\ell^2$ on shell and the bulk term is $\frac{2}{\kappa^2\ell^2}\int_{r_0}^{r(z=\delta)}A\,dr$.  Adding the GHY term $-\frac{1}{\kappa^2}A'(r(z=\delta))$ and the counterterm $\frac{1}{\kappa^2\ell}A(r(z=\delta))$ and taking $\delta\to0$ then gives the result.

    The contribution to the action from the curvature singularity across the constraint surface is the sum of the Gibbons-Hawking-York terms $-\frac{1}{\kappa^2}\int d^2x\sqrt{h}\,K$ for the two sides of the seam (each computed with its own outward normal).  Each side contributes $+\frac{2\pi \beta q}{\kappa^2}\cosh(2r_0/\ell)$,  so the full action is,  
    \begin{equation}
    \begin{aligned}
 S&=\frac{2\pi \beta}{\kappa^2}\big(-q\cosh(2r_0/\ell)
 + 2 q\cosh(2r_0/\ell)\big)+S_{\chi_\infty}\\
 &=\frac{2\pi \beta q}{\kappa^2
 }\cosh(2r_0/\ell)+S_{\chi_\infty}.
\end{aligned}
\end{equation}
    With this we can begin to analyze the action for Dirichlet and Neumann boundary conditions for the axion separately.

\subsubsection{The Dirichlet Action}
    Let us first discuss the case in which we fix Dirichlet boundary conditions for the axion. We fix the axion on the right boundary to be $\chi=\chi_R$ and on the left to be $\chi=\chi_L=-\chi_R$. We can use \eqref{eq:chiRelay} to determine $r_0$ in terms of the fixed boundary value of the axion, finding
    \begin{equation}
    r_0=\ell\operatorname{artanh}\!\left(e^{-\chi_R}\right).
\end{equation}
    Additionally, we can use \eqref{eq:A0Relay} to determine $q$ in terms of $A_0$ and $r_0$,
    \begin{equation}
    q=\frac{\,A_0}{\pi\ell\beta\sinh(2r_0/\ell)}
 =\frac{\,A_0}{\pi \ell\beta}\sinh(\chi_R).
 \label{eq:qdirichlet}
\end{equation}
    Putting these both into the action computed above and recalling that $S_{\chi_\infty}=0$ when using Dirichlet boundary conditions for the axion, we find, 
    \begin{equation}
    S_{CWH}^{(D)}=\frac{2A_0}{\kappa^2\ell}\cosh(\chi_R).
 \label{eq:toroidal_dAction}
\end{equation}
    Note that this  action is linear in the constrained area $A_0$, and thus there are no saddle points for any $\chi_R$ at which $\cosh(\chi_R)\neq 0$.  However, for complex $\chi_R$ there are special values $\chi_R=(2n+1)\frac{i\pi}{2}$ with $n\in\mathbb{Z}$ for which the action becomes identically zero so that \textit{every} complex value of  $A_0$ is a stationary point.

    \subsubsection{The Neumann Action}
    If we instead impose Neumann boundary conditions on the axion by fixing $q$ rather than $\chi_R$, we need to include an additional boundary term in the action in order to maintain a good variational principle. We can obtain this modified action simply by taking a Legendre transform of the action we obtained for the Dirichlet problem. For this we note that in the canonical momentum conjugate to the Dirichlet boundary condition $\chi_R$ is,

    \begin{equation}
    p_{\chi_R}=\frac{\partial S_{CWH}^{(D)}}{\partial\chi_R}
 =2\frac{A_0}{\kappa^2\ell}\sinh(\chi_R)
 =\frac{2\pi \beta q}{\kappa^2}.
\end{equation}
    Using this to take the Legendre transform of the action with Dirichlet boundary conditions we find that, 
    \begin{equation}
    \begin{aligned}
    \label{eq:STN2}
 S_{CWH}^{(N)}&=S_{CWH}^{(D)}-\frac{2\pi \beta q\,\chi_R}{\kappa^2}\\
 &=\frac{2\pi\beta q}{\kappa^2}\left(
 \sqrt{1+\frac{A_0^2}{\pi^2 \ell^2 \beta^2q^2}}
 -\operatorname{arsinh}\bigg(\frac{\pi\ell\beta  q}{\,A_0}\bigg)\right),
 \qquad(q>0).
\end{aligned}
\end{equation}
In parallel with the JT wormholes studied in section \ref{G3wormhole}, this action becomes arbitrarily negative at small real positive $A_0$.  As a result, defining a Neumann partition function $Z_N$ by integrating $e^{-S_{CWH}^{(N)}}$ over $A_0 \in {\mathbb R}^+$ would give a divergent result.   Again, the existence of a nowhere-vanishing hypersurface orthogonal Killing field means that $S_{CWH}^{(N)} = \beta E$, where $E$ is the energy computed in the Wick-rotated Lorentzian constrained wormhole, so that the above divergence is a direct result of the energy becoming 
arbitrarily negative for small $A_0$.

As discussed in the introduction, the fact that the divergence arises at small $A_0$ makes it UV-sensitive.  However, our goal here is merely to investigate the dependence of wormhole contributions on boundary conditions in toy models and not to study the microscopic dynamics in detail.  We will therefore follow the procedure outlined in the introduction of first defining the partition function $Z_N(q)$ at imaginary values of $q$. 

Some of the discussion from section \ref{G3wormhole} will again apply to this toroidal case, though there are significant differences.  In particular, while the integral of $e^{-S_{CWH}^{(N)}}$ over $A_0 \in {\mathbb R}^+$ will converge at the origin for imaginary $q$, the behavior is power-law in $A_0$, with a power that depends on $q$.  Deforming the direction in which the $A_0$-contour approaches the origin is thus not helpful in maintaining convergence when analytically continuing $q$ away from imaginary values.

In addition, in considering imaginary $q$ we immediately encounter the other issue foreshadowed in the introduction, in that the Neumann constrained-wormhole action \eqref{eq:STN2} has branch points at $A_0/\pi\ell\beta q = \pm i$, so that for imaginary $q$ there is a branch point on the positive-real $A_0$-axis.  We therefore fall back on the procedure suggested in the introduction of simply defining $Z_N$ at complex $q$ in terms of the Fourier transform of $Z_D.$

\subsection{The Partition Functions}

Let us thus begin by computing the partition function for the Dirichlet boundary problem. As noted above, doing so require a choice of measure for the $A_0$-integral.  For simplicity, and in the spirit of continuing to treat our system as a toy model, we will simply take this  measure to be $dA_0$ (though this choice is in  no way preferred by the analysis we have performed), so that we have 
\begin{equation}
    Z_D[\chi_R]=\int_0^\infty dA_0\,e^{-S_{CWH}^{(D)}[A_0,\chi_R]}
    =\int_0^\infty dA_0\,e^{-\frac{2A_0}{\kappa^2\ell}\cosh(\chi_R)}
 =\frac{\kappa^2\ell}{2\cosh(\chi_R)}.
\end{equation}

As discussed in section \ref{G3wormhole}, with this choice of measure the Fourier transform of $Z_D$ is {\it not} in fact expected to exhibit a logarithmic singularity at $q=0$ as, if one performs the Fourier transform by integrating over $\chi_R$ (say, using the saddle-point approximation) before integrating over $A_0$, then the associated one-loop factor generates an extra factor of $A_0^{-1/2}$ that causes the  saddles of the $A_0$-integral to appear in oppositely-signed pairs.  The  winding effects associated with the two members of each pair then simply cancel against each other each time we circle the origin of the complex $q$-plane. 

Indeed, with the above choices we simply write $Q=-iq$ to find
\begin{equation}
    \begin{aligned}
    Z_N[q]&=\frac{1}{\sqrt{2\pi}}\int_{-\infty}^{+\infty}d\chi_R\,
    \frac{\kappa^2\ell}{2\cosh(\chi_R)}
    \exp\!\left(2\pi i
    \beta\chi_R Q/\kappa^2 \right)\\
    &=\sqrt{\frac{\pi}{2}}\, \frac{\kappa^2\ell}
    {2\cosh \!\left(\frac{\pi^2 \beta  Q}{\kappa^2}\right)}=\sqrt{\frac{\pi}{2}}\, \frac{\kappa^2\ell}
    {2 \cos \!\left(\frac{\pi^2  \beta q}{\kappa^2}\right)},
    \label{eq:ZNTF}
    \end{aligned}
\end{equation}
where the transition to the final line is a standard exercise in complex analysis integral that can be computed using the fact that $\cosh(x+i\pi)=-\cosh(x)$ to define a useful closed contour over which the integral can be expressed in terms of residues  of the enclosed poles.  As expected, \eqref{eq:ZNTF} shows that $Z_N$ has no logarithmic singularity.

It is also interesting to evaluate 
\begin{equation}
    \begin{aligned}
    Z_N[q]&=\frac{1}{\sqrt{2\pi}}\int_{-\infty}^{+\infty}d\chi_R\,
    Z_D[\chi_R] 
    \exp\!\left(2\pi i \beta \chi_R Q/\kappa^2\right)\\
    &=
    \frac{1}{\sqrt{2\pi}}\int_{-\infty}^{+\infty}d\chi_R\,
\int_0^\infty dA_0\,e^{-2\frac{A_0}{\kappa^2\ell}\cosh(\chi_R)}    \exp\!\left(2\pi i\beta\chi_R Q/\kappa^2\right)
    \label{eq:ZNTF2}
    \end{aligned}
\end{equation}
by first performing the integral over $\chi_R$ (for real $Q>0$) and only later integrating over $A_0.$  The $\chi_R$ integral can be performed in the stationary phase approximation, though in this case it is simpler to recover the asymptotic expansion by using Bessel-function identities to write 
\begin{align} 
&f(Q,A_0) :=   \frac{1}{\sqrt{2\pi}}\int_{-\infty}^{+\infty}d\chi_R\,
e^{-2\frac{A_0}{\kappa^2\ell}\cosh(\chi_R)}    \exp\!\left(2\pi i\beta\chi_R Q/\kappa^2\right)
= \sqrt{\frac{2}{\pi}}\, K_{2\pi i\beta Q/\kappa^2}\left(\frac{2A_0}{\kappa^2\ell}\right) \nonumber \\
&\approx\;
\begin{cases}
\dfrac{\kappa}{\sqrt2\left(\frac{A_0^2}{\ell^2}-\pi^2\beta^2Q^2\right)^{1/4}}\;
\exp\!\left[-\dfrac{2}{\kappa^2}\left(\sqrt{\frac{A_0^2}{\ell^2}-\pi^2\beta^2Q^2} + \pi\beta Q\arcsin\dfrac{\pi\ell\beta Q}{A_0}\right)\right],
& 0< \pi\ell\beta Q < A_0, \\[3ex]
\dfrac{\sqrt2\,\kappa e^{-\pi^2\beta Q/\kappa^2}}{\left(\pi^2\beta^2Q^2-\frac{A_0^2}{\ell^2}\right)^{1/4}}\; 
\cos\!\left[\dfrac{2}{\kappa^2}\left(\pi\beta Q\,\operatorname{arccosh}\dfrac{\pi\ell\beta Q}{A_0} - \sqrt{\pi^2\beta^2Q^2-\frac{A_0^2}{\ell^2}}\right) - \dfrac{\pi}{4}\right],
& \pi\ell\beta Q > A_0 >0.
\end{cases}
\label{eq:Ktrans}
\end{align}

The final results then show that for $\pi \beta Q
\ell < A_0$ the leading-order saddle-point result is  $f \approx e^{-S_{CWH}^{(N)}}$ with $S_{CWH}^{(N)}$ given by \eqref{eq:STN2} evaluated using the branch of the square root that makes this amplitude decay, as one would expect.  In particular, the relevant branch is the one associated with what were called short wormholes in \cite{AxionWormholes}, in agreement with the predictions of that reference.  

In contrast, for $\pi \beta Q
\ell > A_0$, we must sum over the two branches of this square root.  This can also be seen from the fact that $Z_D$ is even in $\chi_R$, and that the Fourier transform of an even function is real, so saddles must appear in complex-conjugate pairs.  In both cases, however, we should include only the 
principal branch of ${\rm arcsin} \left(\frac{\pi\ell\beta Q}{A_0} \right) = -i\,{\rm arcsinh} \left(\frac{\pi\ell\beta q}{A_0} \right)$.

If we now use the above result to compute
\begin{equation}
\label{eq:ANA0int}
    Z_N(iQ):= \int_{{\mathbb R}^+} dA_0 \, f(Q, A_0),  \end{equation}
we may again use the stationary phase approximation.  However, it is manifest that the only stationary point with $A_0 \in {\mathbb R}^+$ is at $A_{crit} = \pi \beta Q\ell$, near which $S_{CWH}^{(N)}$ behaves as $(A_0 - A_{crit})^{3/2}$; i.e., this is not a regular saddle point.  Nonetheless, it dominates in the stationary phase approximation as can be seen from the fact that it gives $e^{-S_{CWH}^{(N)}} \sim e^{-\pi^2\beta Q/\kappa^2}$, which agrees with the asymptotic behavior of \eqref{eq:ZNTF}.  There is also a similar stationary point at $-A_{crit}$.

\begin{figure}[h!]
\centering
\includegraphics[width=.75\linewidth]{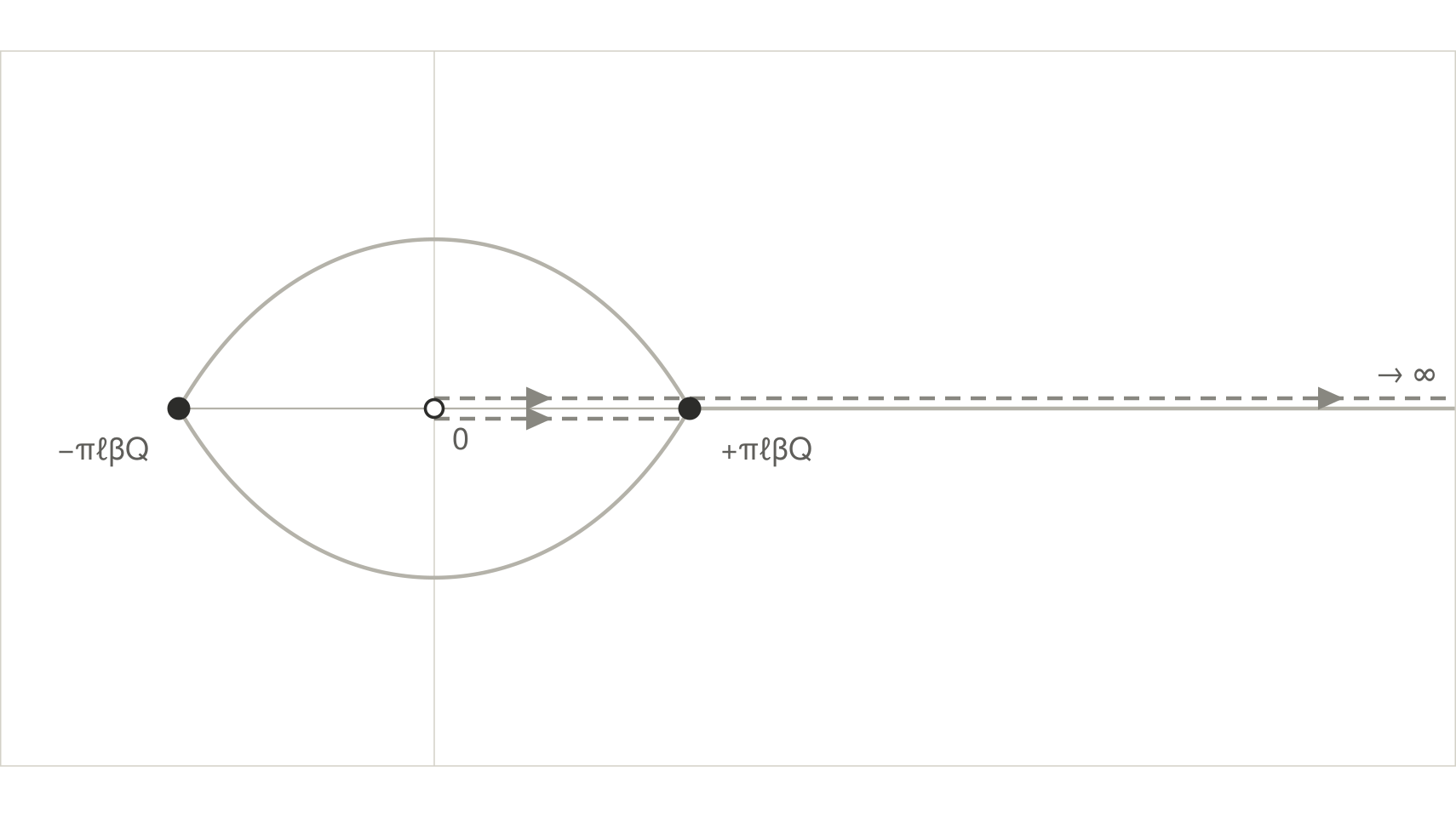}
\caption{ For real $Q=-iq$, the integral \eqref{eq:ANA0int} is initially defined by integrating along the dashed grey contour.  The two pieces of this contour for $A_0 < A_{crit} = \pi\ell\beta Q$ correspond to the two branches of $S_{CWH}^{(N)}$ over which we must sum.  These contours can be deformed to a pair of descent contours through $A_{crit}$, with one circling the origin along the grey curves infinitely many times in the clockwise direction, and with the other doing so in the counter-clockwise direction.  These curves in fact each pass through an infinite family of additional critical points, with two such points on each sheet of the Riemann surface; see the associated comments in footnote \ref{foot:StokesInf}.  We note that near the points $ \pm A_{crit}$, we have $S_{CWH}^{(N)} \sim (A_0 \mp A_{crit})^{3/2}$.  As a result, $A_0 = \pm A_{crit}$ are branch points at which the descent contour changes direction discontinuously as shown.}
\label{fig:A0}
\end{figure}

Indeed, for $A_0 \ge A_{crit}$, the original integration contour coincides with a steepest descent contour through $A_{crit}$.  We may then deform the integration contour for $A_0 \le A_{crit}$ to a set of descent contours as well.  In doing so, we should recall that for $A_0 \le A_{crit}$ the integrand is a sum of two branches, and that each branch should lead to an independent set of descent contours.  Indeed, for one branch this segment should be deformed to a descent contour that winds clockwise  infinitely many times around the logarithmic branch point at the origin, and along which the amplitude decays with each circuit of the origin.  For the other branch the segment $A_0< A_{crit}$ should instead be deformed to an analogous contour that winds counter-clockwise around the origin (but which, for that branch, again gives decay along the contour).  See figure \ref{fig:A0} for the projections of these contours onto the initial sheet of the associated Riemann surface for $A_0.$  As can be seen in the figure, this descent contour in fact passes through an infinite collection of critical points (each of which is a branch point).\footnote{\label{foot:StokesInf} There are thus in some sense an infinite number of Stokes' transitions taking place simultaneously at real values of $Q=-iq$.  However, a careful analysis shows that no intersection numbers actually change across this Stokes' line.  This is a consequence of each branch point being connected to the next by {\it two} of its three ascent/descent curves at real $Q$.  One can also see this from the fact that the sum over branch-point contributions in fact converges to $Z_N$ as discussed below.}   Summing the resulting contributions reproduces the expansion of \eqref{eq:ZNTF} in terms of powers of $e^{-\pi^2\beta Q/\kappa^2}$ for real $Q$.

Comparing with \eqref{eq:AT2}, we see that each of the above critical points corresponds to a spacetime wormhole, though for $A_0 = -\pi\beta Q \ell$ the wormhole is not real Euclidean. 
All of these wormholes thus contribute to our $Z_N$ for imaginary $q$.  

An interesting property of the current system is that $S_{CWH}^{(D)}$ can be stationary with respect to $A_0$ only at the discrete set of values $\chi_R = i\left( \frac{\pi}{2} + \pi n\right)$ (with $n \in {\mathbb Z}$) for which $\cosh(\chi_R)=0$.  Thus all wormholes take one of these values of $\chi_R$.  Thus while $Z_D$ receives no wormhole contributions at other values of $\chi_R$ (where it is given by a single endpoint contribution alone), one may say that the poles at the above values of $\chi_R$ come from integrating over the non-compact moduli space of wormholes for each such $\chi_R$, though since $Z_D$ integrates only over  $A_0 \in {\mathbb R}^+$ the relevant wormholes are only those with $A_0>0$.  In contrast, as shown in figure \ref{fig:A0}, wormholes with $A_0<0$ do contribute to $Z_N$ at imaginary $q$.  So this is again an example of the phenomenon that motivated this work, and we have identified another example where the Dirichlet and Neumann partition functions receive contributions from distinct sets of wormholes.  


\section{Asymptotically Flat Axion Wormholes}\label{sec:flat wormholes}
\label{sec:asFlat}

Let us now perform a corresponding  analysis of $d$-dimensional asymptotically flat axion wormholes for general $d > 2$.  We again consider both Dirichlet and Neumann boundary conditions for the axion $\chi$.  In Euclidean signature we have
\begin{align}
    S &= S\sub{EH} + S_\chi\,, \label{eq:action}\\
    S\sub{EH} &= -\frac1{2\kappa^2}\int_\MM\df^d x\sqrt{g}R - \frac1{\kappa^2}\int_\pMM \df^{d-1}x\sqrt{h}K\,,\\
    S_\chi &= \frac1{4\kappa^2} \int_\MM\df^d x\sqrt{g}\partial^\mu\chi\partial_\mu\chi & &\text{(Dirichlet)}\\
    S_\chi &= \frac1{4\kappa^2} \int_\MM\df^d x\sqrt{g}\partial^\mu\chi\partial_\mu\chi - \frac1{2\kappa^2}\int_\pMM\df^{d-1}x\sqrt h \chi n^\mu\partial_\mu\chi & &\text{(Neumann),}
\end{align}
where $\kappa^2 = 8\pi G_N$ and where we have used a gravitational normalization involving $\kappa^2$ for the scalar-field action. We impose spherically symmetry and thus require 
\begin{align}
    ds\sub E^2 &= \df r^2 + a^2(r)d\Omega_{d-1}^2 := \df r^2 + g_{ij}\df x^i \df x^j\,, & \chi &= \chi(r)\,, \label{eq:ansatz}
\end{align}
where $d\Omega_{d-1}^2 = a^{-2}g_{ij}\df x^i \df x^j$ is the round metric on a unit $S^{d-1}$ and where we will later define $A(r)$ to be the actual area of a cross-section (including the required factor of the volume of the unit $S^{d-1}$).

In this case, if we require our wormholes to be smooth at infinity, there is no freedom to introduce a parameter $\beta$.  Instead, the periods of the various coordinates are fixed by the requirement that there be no conical singularities extending to infinity.  If we think of the azimuthal angle as Euclidean time, we might therefore say that here we simply consider the particular case $\beta=2\pi$.

The Gibbons-Hawking-York term of a $d$ dimensional ball of radius $r$ is proportional to $r^{d-2}$.  As a result, the above  action diverges as $r\to\infty$. Following \cite{Hawking:1996}, we take the renormalized action to be
\begin{align}
    S_{\rm ren} := S - S^{(0)}\,,
\end{align}
where $S^{(0)}$ is the action of a reference vacuum solution $\mathcal M_0$ with boundary $\partial\mathcal M_0 = \partial \mathcal M$.

In the Dirichlet case we as usual take $\chi$ to have boundary values $\chi_L, \chi_R$, and define $k = \chi_R - \chi_L$. In the Neumann case, we take $n^\mu\partial_\mu \chi$ to be constant across each boundary, with magnitude specified by the conserved axion flux 
\begin{align}
    q :={}& \frac1{V_{d-1}}\int \df^{d-1}x \sqrt{h}\, n^\mu_R \partial_\mu \chi\,. \label{eq:flat q def}
\end{align}
Here
\begin{align}
    V_{d-1} = \frac{2\pi^{d/2}}{\Gamma\big(\frac d2\big)}
\end{align}
is the volume of a unit $S^{d-1}$.

By convention we define $q$ using the normal $n_R$ that points toward the right boundary. As stated in the introduction, we take 
the Euclidean flux $q = iQ$ to be imaginary on the defining contour. The Neumann boundary term is then $-iV_{d-1}kQ/2\kappa^2$ and, in particular, we take the Neumann and Dirichlet partition functions to be related by a Fourier transform.

As usual, to find wormhole saddles we must insert constraints.
We choose to constrain the area $A_0$ of the surface with $\chi = \chi_0$ for some fixed $\chi_0$, so that the constrained action becomes
\begin{align}
    S(A_0) &= S\sub{EH} + S_\chi + \lambda_A\left[\int\df^dx\sqrt{g}n^\mu\partial_\mu\theta(\chi - \chi_0) - A_0\right], \label{eq:constrained action}
\end{align}
where $n^\mu = \partial^\mu\chi/\sqrt{\partial_\nu\chi\partial^\nu\chi}$ is a unit normal to the $\chi_0$ surface whose sign depends on the sign of the boundary condition set by $k$ or $q$. 
For simplicity we again assume $\chi_R = -\chi_L$ and then take $\chi_0=0$.

\subsection{Classical solutions}
The analysis of the equations of motion is standard, but for completeness we now include a brief review.  In the absence of the constraint term, the equations of motion are
\begin{align}
    0 &= \frac1{\sqrt{g}}\vderiv S{\chi} = -\frac1{2\kappa^2}\nabla^2\chi \label{eq:chi eom} \\
    0 &= \frac2{\sqrt{g}}\vderiv S{g^{\mu\nu}} = -\frac1{\kappa^2}\left(R_{\mu\nu} - \frac12 Rg_{\mu\nu} - \frac12 T_{\mu\nu}\right) \label{eq:g eom}\,,
\end{align}
where
\begin{align}
    T_{\mu\nu} = \frac{4\kappa^2}{\sqrt{g}}\vderiv{S_\chi}{g^{\mu\nu}} = \partial_\mu\chi\partial_\nu\chi - \frac12 g_{\mu\nu}\partial_\rho\chi\partial^\rho\chi\,.
\end{align}
With our ansatz, (\ref{eq:chi eom}) becomes
\begin{align}
    0 = \frac1{a^{d-1}}\partial_r(a^{d-1}\partial_r\chi)\,. \label{eq:spherical chi}
\end{align}
The trace of (\ref{eq:g eom}) is $R = \frac12 (\partial \chi)^2$, with which (\ref{eq:g eom}) reduces to
\begin{align}
    R_{\mu\nu} = \frac12\partial_\mu\chi\partial_\nu\chi\,.\label{eq:simplified Einstein}
\end{align}
More explicitly, we find 
\begin{align}
    \frac12 (\partial_r\chi)^2 = R_{rr} &= -(d-1)\frac{\partial_r^2 a}a, \ \ \ {\rm and} \label{eq:spherical grr}\\
    0 = R_{ij} &= \partial_r\Gamma_{ij}^r + 2\Gamma_{j[i}^r\Gamma_{k]r}^k - \Gamma_{jr}^k\Gamma_{ik}^r + \tilde R_{ij}\nonumber \\
    &= \left[(d-2)\frac{1-(\partial_r a)^2}{a^2} - \frac{\partial_r^2 a}{a}\right]g_{ij}\,,\label{eq:spherical gij}
\end{align}
where $\tilde R_{ijkl}$ is the Riemann tensor of a $(d-1)$-sphere, $\tilde R_{ijkl} = \frac1{a^2}(g_{ik}g_{jl} -  g_{il}g_{jk})$. 

Integrating (\ref{eq:spherical chi}) yields
\begin{align}
    \partial_r\chi = \frac{c_1}{a^{d-1}}\,,\label{eq:dchi of a}
\end{align}
and combining equations (\ref{eq:spherical grr}-\ref{eq:dchi of a}) we find
\begin{align}
    -\frac{c_1^2}{a^{2(d-2)}} = 2(d-1)(d-2)[1-(\partial_ra)^2]\,.
\end{align}
Inserting \eqref{eq:dchi of a} into \eqref{eq:flat q def}, we see $c_1 = q$. Treating $a$ as a new radial coordinate, the metric becomes
\begin{align}
    ds^2 = \frac{\df a^2}{1 + \frac{q^2/2(d-1)(d-2)}{a^{2(d-2)}}} + a^2d\Omega_{d-1}^2 \label{eq:a metric}
\end{align}
and the axion gradient takes the form
\begin{align}
    \partial_a\chi = \pm\frac{q}{a^{d-1}}\frac{1}{\sqrt{1 + \frac{q^2/2(d-1)(d-2)}{a^{2(d-2)}}}}\,.\label{eq:a dchi}
\end{align}

For real $q$, the metric function $a$ described by (\ref{eq:a metric}) has no minimum, and the manifold it describes has only one asymptotic region. Noting that adding the constraint term $S_{con}$ allows the extrinsic curvature to be discontinuous across the surface $\chi=\chi_0$, we now construct constrained wormhole saddles by gluing together two such solutions along surfaces of constant $\chi$; it is shown in \cite{Held:2026bbo} that for $\chi_0 = \frac12(\chi_L + \chi_R)$, the glued wormhole indeed satisfies the constrained equations of motion on the $\chi_0$ surface. 

However, as discussed in the introduction we will in fact focus on the case of imaginary $q$.  With such boundary conditions the metric can indeed have a minimum, and there are saddles satisfying all equations of motion.  These are just the familiar axion wormholes of \cite{Giddings:Original_Axion_Wormhole}.  However, to understand precisely how such wormholes contribute to the Neumann partition function $Z_N$ we will still proceed by expressing $Z_N$ as an integral over constrained wormholes.  

\subsection{The action}
The trace $R = \frac12(\partial\chi)^2$ of Einstein's equation (\ref{eq:g eom}) implies that the bulk action vanishes away from the $\chi_0$ surface. Furthermore,  $\chi$ is continuous across this surface.  As a result, the only contribution of this surface to the action is the Gibbons-Hawking-York term resulting from cutting the wormhole along this surface. The action is then
\begin{align}
    S_{\rm ren} = \lim_{a_\partial\to\infty}2\left(S\sub{GHY}(a_\partial) + S\sub{GHY}(a_0) + S_{\chi,\partial}(a_\partial) - S^{(0)}\sub{GHY}(a_\partial)\right),
\end{align}
where $a_0 := (A_0/V_{d-1})^{\frac1{d-1}}$ is the scale factor of the surface where $\chi=\chi_0$.  In addition, the axion boundary term is
\begin{align}
    S_{\chi,\partial}^{(D)}(a_\partial) &= 0 & &(\text{Dirichlet}) \\
    S_{\chi,\partial}^{(N)}(a_\partial) &= -\frac1{2\kappa^2}\int d^{d-1}x\sqrt{h}\,\chi n^\mu\partial_\mu\chi = -\frac{V_{d-1}}{2\kappa^2}\chi(a_\partial)q & &(\text{Neumann})\,,
\end{align}
and the Gibbons-Hawking-York term of a constant $a$ surface is
\begin{align}
    S_{\text{GHY}}(a) &= -\frac1{\kappa^2}\int \df^{d-1}x\;\sqrt{h}\,K \nonumber \\
     &= -\frac1{\kappa^2}\int \df^{d-1}x\;\sqrt{h}\,\sign(n_a)\frac{d-1}{a\sqrt{g_{aa}}}  \nonumber \\
    &= -\sign(n_a)(d-1)\frac{V_{d-1}}{\kappa^2}\sqrt{a^{2(d-2)} + \frac{q^2}{2(d-1)(d-2)} }
\end{align}
in the above classical solution, while
\begin{align}
    S\sub{GHY}^{(0)}(a) &= -\sign(n_a)(d-1)\frac{V_{d-1}}{\kappa^2} a^{d-2}
\end{align}
in the reference vacuum solution. The difference between the two boundary Gibbons-Hawking-York boundary terms is $O\left(a^{-(d-2)}\right)$, so we have
\begin{align}
    S_{CWH}^{(D)} &= \frac{2(d-1)}{\kappa^2}V_{d-1}a_0^{d-2} \sqrt{1 + \psi^2} & &(\text{Dirichlet}) \label{eq:dirichlet action}\\
    S_{CWH}^{(N)} &= \frac{2(d-1)}{\kappa^2}V_{d-1}a_0^{d-2}\sqrt{1 + \psi^2} - \frac{V_{d-1}}{2\kappa^2}kq & &(\text{Neumann})\,,\label{eq:neumann action}
\end{align}
where 
\begin{align}
    \psi :={}& \frac{q/a_0^{d-2}}{\sqrt{2(d-1)(d-2)}}\,.
\end{align}
The Dirichlet and Neumann boundary conditions are related through
\begin{align}
    k &= 2\int_{a_0}^\infty \df a\;\partial_a\chi = 2q\int_{a_0}^\infty \frac{\df a}{a^{d-1}}\left(1 + \frac{1}{2(d-1)(d-2)}\frac{q^2}{a^{2(d-2)}}\right)^{-\frac12} \nonumber \\
    &= 2\sqrt{2\alpha_d}\sinh^{-1}\psi,
\end{align}
where $\alpha_d := \frac{d-1}{d-2}$ and $k$ is given by the branch of $\sinh^{-1}\psi$ that is real for real $\psi$. 

\subsection{The partition function}

As in previous sections, the Dirichlet and Neumann wormhole partition functions are
\begin{equation}
    Z_{D/N}     \approx \int_0^\infty\df A_0\;e^{-S_{CWH}^{D/N}(A_0)}\,, \label{eq:partition from F}
\end{equation}
where our use of $\approx$ indicates that the precise measure is undetermined at the order in $\kappa$ at which we work. From the above we have
\begin{align}
    S_{CWH}^{(D)}(A_0) &= \frac{2(d-1)}{\kappa^2}V_{d-1}a_0^{d-2}\cosh\left(\frac k{2\sqrt{2\alpha_d}}\right) \\
    S_{CWH}^{(N)}(A_0) &= \sqrt{2\alpha_d} \frac{V_{d-1}}{\kappa^2}q\left[\frac1\psi \sqrt{1 + \psi^2} - \sinh^{-1}\psi\right].
    \label{eq:AsFSDN}
\end{align} 
It is convenient to change variables to $B = a_0^{d-2}$ and, again in the spirit of treating our systems as toy models, to simply define our partition functions using the measure  $\df B$. The Dirichlet partition function is then
\begin{align}
    Z_D(k) &:= \frac{\kappa^2}{2(d-1)V_{d-1}}\sech\left(\frac k{2\sqrt{2\alpha_d}}\right)
\end{align}
In direct parallel with the discussion at the end of section \ref{sec:tori}, 
defining the Neumann partition function at $q=iQ$ to be the (appropriately normalized) Fourier transform $\tilde Z_D(Q)$ of $Z_D$, we find
\begin{align}
    Z_N(iQ) &:= \int_{-\infty}^\infty \df k\, e^{i\frac{V_{d-1}}{2\kappa^2}kQ} Z_{D}(k)
    = \frac{\pi\kappa^2\sqrt{2\alpha_d}\,}{(d-1)V_{d-1}}\,
      \sech\left(\frac{\pi \sqrt{2\alpha_d}\,V_{d-1}Q}{2\kappa^2}\right) \nonumber \\
    &= \frac{\pi\kappa^2\sqrt{2\alpha_d}\,}{(d-1)V_{d-1}}\,
      \sec\left(\frac{\pi \sqrt{2\alpha_d}\,V_{d-1}q}{2\kappa^2}\right),
    \label{eq:Fourier Neumann}
\end{align}
where as usual we perform the above integral at real $Q=-iq$ and then analytically continue.

The discussion of saddles is then manifestly identical to that of section \ref{sec:tori}.  In particular, just as in that case, one may instead Fourier transform the {\it integrand} of the Dirichlet partition function to define a set of branches of
$e^{-S_{CWH}^{(N)}}$ over which one should sum in order  to calculate the above $Z_N$ as $\int_{B \in {\mathbb R}^+} dB e^{-S_{CWH}^{(N)}}$. In particular, we see that, up to the insertion of constant factors, our current expressions for $Z_N$, $Z_D$ as integrals over $B=a_0^{d-2}$  take precisely the same form as the corresponding integrals over $A_0$ in section \ref{sec:tori}. We thus again refer the reader to figure \ref{fig:A0} for an illustration of how our recipe for constructing $Z_N$ translates to a set of descent contours on the Riemann surface for complex $B$ (with the understanding that the critical points on the $B$ plane are $\pm Q/\sqrt{2(d-1)(d-2)}$ rather than $\pm\pi\ell\beta Q$). We also remind the reader that this figure shows explicitly that for imaginary $q$ the Neumann partition function $Z_N$ receives contributions from wormholes with $B<0$, while such wormholes do not contribute to $Z_D$ for e.g. $\chi_R = i\pi\sqrt{\alpha_d/2}$ or, equivalently, for $k=i\pi\sqrt{2\alpha_d}$ (at least when approached from values with smaller imaginary parts).

\section{Discussion}
\label{sec:Disc}

The above work investigated Dirichlet and Neumann wormhole partition functions in low-energy effective theories of gravity for JT gravity coupled to a scalar field, for asymptotically-AdS scalar/axion wormholes with toroidal AdS boundaries, and for asymptotically flat scalar/axion wormholes.  The systems were treated at leading order in the semiclassical expansion.  Only certain constrained-wormhole degrees of freedom were treated in detail.  In particular, we
first constrained the cross-sectional area $A_0$  on the surface where the axion/scalar took a certain value, after which we computed the Euclidean actions $S_{CWH}^{(D)},S_{CWH}^{(N)}$ of the corresponding constrained saddles defined by Dirichlet and Neumann boundary conditions.  The Dirichlet and Neumann partition functions $Z_D$ and $Z_N$ were then defined as integrals of $e^{-{S_{CWH}^{(D)}}}$ and $e^{-{S_{CWH}^{(N)}}}$ over $A_0$.   

In addition, we saw explicitly that the final results could depend on the choice of measure for these $A_0$-integrals, though such one-loop corrections are beyond the scope of this work.  Our treatments thus have the status of toy models, illustrating possible behaviors that can arise.  In particular, in practice we simply postulated relevant one-loop features and then explored the implications of those choices.

Nevertheless, our toy models support the suggestion of \cite{AxionWormholes} that the relevance of wormhole saddles to partition functions can depend strongly on the choice of boundary conditions, which here amounted to the choice of a Neumann condition $q$ or a Dirichlet condition $k$ (or $\chi_R$).  This was particularly clear in the case of JT gravity coupled to a scalar for which the Dirichlet partition function $Z_D$ receives contributions from wormhole saddles when $\arg{k} = -\pi/3$, which occurs at real $q$  for the upper right saddle ($A_0^0$) in  figure \ref{fig:JTqreal} whose ascent curve  has zero intersection number with the contour of integration for $Z_N$.    

An example in the other direction is then given by studying this same system  at negative imaginary $q$.  In that case one can see from the bottom left panel in figure \ref{fig:windingcontours} that the ascent curves for the two left saddles both flow to the origin, and thus have intersection number $\pm 1$ with the contour of integration for $Z_N$.  So these saddles contribute to $Z_N$ (regardless of the possible presence of an extra factor of $A_0^{-1/2}$ in the integrand of $Z_N$).   But \eqref{eq:JTqk} then gives $\arg (k)=\pi/6, 5\pi/6$ so $|{\rm Re}\, k|> |{\rm Im}\, k|$ and these wormholes do not contribute to $Z_D$.  Indeed, it turns out that (again regardless of the presence of an extra factor of $A_0^{-1/2}$), the wormhole contributions to $Z_D$ and $Z_N$ agreed {\it only} for $\arg (q)$ within $\pi/4$ of the positive imaginary axis (and on the principal sheet when there is a logarithmic singularity). We also saw that there is a sense in which a similar phenomenon occurs for the toroidal-AdS and asymptotically flat axion wormholes of section \ref{sec:tori} and \ref{sec:asFlat}.

For all of our systems, the use of real Neumann boundary conditions resulted in the energies of our constrained wormholes being unbounded below.    As a result, we chose to define $Z_N$ as an integral over $A_0$ computed at imaginary values of the Neumann boundary condition and to then define $Z_N$ more generally by analytic continuation.  However, doing so led to the presence of branch points on the integration contour.  In order to choose a definite set of branches, we thus in fact defined $Z_N$ at imaginary Neumann boundary conditions in terms of the Fourier transform of $Z_D.$  It would be interesting to understand if there are other families of constrained wormholes for which the energies with real Neumann boundary conditions remain bounded below (and which thus provide a more direct definition of $Z_N$), or whether this behavior is somehow generic.

In investigating the JT + scalar model in section \ref{G3wormhole}, we also found an interesting potential logarithmic singularity in $Z_N$ whose presence or absence depends on the character of one-loop corrections.  This potential singularity is associated with the winding of integration contours for $A_0$ as one analytically continues the phase of the Neumann boundary condition $q$.  When the logarithmic singularity does arise, this winding can be used to identify additional cases where wormholes contribute to $Z_N$ without contributing to $Z_D$.

The toroidal-AdS wormholes of section \ref{sec:tori} and the asymptotically-flat wormholes of section \ref{sec:asFlat} did not exhibit this winding behavior.  However, these cases brought two additional subtleties of their own.  The first is that, as is well known,  for on-shell wormholes the asymptotic values of the axion/scalar are in fact independent of the flux through the wormhole.  The second is that, in our formalism, the on-shell wormholes do not correspond to smooth saddles in $A_0$.  Instead, the relevant Neumann actions behave like $(A_0-A_{crit})^{3/2}$ near the relevant critical points $A_{crit}.$  This was also the case for the spherical asymptotically-AdS axion wormholes studied in \cite{AxionWormholes}.

On a final note,  it is interesting to recall that the structures explored here and in \cite{AxionWormholes,Held:2026bbo} were not seen in previous studies of `stability' of Euclidean axion wormholes \cite{Loges:2022nuw,andriolo:Axion_wormholes_massive_dilaton,Loges:2023ypl,Hertog:Axion-Saxion_Wormholes_Stable,aguilar:Axion_dS_Wormholes,loveridge:Axion_Wormhole_Alternate_Topologies,Marolf:2025evo} that followed the constrain-first approach\footnote{Note that \cite{Loges:2022nuw} also considered a constrained wormhole approach (based on fixing the value of the radial lapse) that did see much of this structure.} of \cite{Garriga:1997wz,Gratton:1999ya,Kol:2006ga}.    Indeed, in the asymptotically flat cases, with Neumann boundary conditions such analyses seemed to find all spherically-symmetric constraint-preserving perturbations of on-shell  axion wormholes to be pure gauge.  On the other hand, section \ref{sec:asFlat}  studied a family of spherical off-shell wormholes whose Neumann action $S_{CWH}^{(N)}$ generally differs from that of the on-shell wormhole despite the fact that our family manifestly satisfies the constraints imposed in \cite{Loges:2022nuw,andriolo:Axion_wormholes_massive_dilaton,Loges:2023ypl,Hertog:Axion-Saxion_Wormholes_Stable,aguilar:Axion_dS_Wormholes,loveridge:Axion_Wormhole_Alternate_Topologies,Marolf:2025evo}.  

The point here appears to be that our family is simply invisible to the gauge-invariant variables used in these references.  Indeed, since our constrained wormholes are built by cutting and gluing pieces of on-shell wormholes, no local gauge-invariant variable on either side of the seam could possibly detect our family of deformations.  The gauge-invariant characterizations of our Neumann family are instead associated with the change $\delta L$ in the total length of the wormhole,  the change $\delta A_0$ in the size of the throat, and the jump $\Delta K$ in the extrinsic curvature at the seam (with the capital $\Delta$ here denoting a jump while the lower-case $\delta$ denotes a change relative to the on-shell wormhole\footnote{Since $\Delta K=0$ for the smooth wormhole, $\delta \Delta K = \Delta K$ and we use the latter notation to prevent clutter.}).  A short calculation (which in fact also applies in both our toroidal-AdS case and the spherical-AdS case of \cite{AxionWormholes}) then gives $\delta L \sim \Delta K \sim \sqrt{\delta A_0}$.  This indicates that the smooth parameters along our family are in fact $\delta L$ and $\Delta K$, so that the above relations are better written as 
\begin{equation}
\delta A_0 \sim \delta L^2 \sim \Delta K^2.
\end{equation}
It is thus clear that our $\delta A_0$ will be invisible at the level of linear fluctuations analyzed in \cite{Loges:2022nuw,andriolo:Axion_wormholes_massive_dilaton,Loges:2023ypl,Hertog:Axion-Saxion_Wormholes_Stable,aguilar:Axion_dS_Wormholes,loveridge:Axion_Wormhole_Alternate_Topologies,Marolf:2025evo}.  We may then also note that, in retrospect, none of their gauge-invariant variables are sensitive to either the change in length $\delta L$ or the jump in extrinsic curvature $\Delta K$; i.e., their variables do not encode the information required to fully reconstruct the geometry when it fails to be smooth.  

However, as discussed above, the toroidal-AdS, spherical-AdS, and asymptotically flat cases satisfy 
$\delta S_{CWH}^{(N)} \sim \delta A_0^{3/2}$, so that we find
$\delta S_{CWH}^{(N)} \sim \delta L^{3}$.  As a result, $\delta L$ does {\it not} contribute to the action at the quadratic order studied in the above references. In this sense it is self-consistent that their treatment does not see this family of deformations.  But the non-trivial dependence at cubic order means that $\delta L$ is nevertheless a modulus that deserves study; it is not a pure-gauge direction to be removed from the path integral.

We thus conclude that the constrain-first approach should generally at least be supplemented by a study of such moduli.  Of course, one would also like to better understand the extent to which the constrain-first approach can be 
justified from first principles. As noted in e.g. \cite{Horowitz:2025zpx}, such a justification must necessarily be tied to a proposal for resolving the so-called conformal factor problem \cite{Gibbons:1978ac} and, presumably, for prescribing an appropriate contour along which the gravitational path integral is to be defined.  It is thus natural to further consider Lorentzian approaches in this regard (see again comments in the introduction).  Some remarks on implications for the constrain-first recipe  will appear in forthcoming work \cite{AKM}.


\acknowledgments
DM thanks Stefano Antonini and Maciej Kolanowski for related discussions.
The work of DM was supported by  NSF grant PHY-2408110,  and by funds from the University of California. ZW was supported by the DOE award
number DE-SC0015655, and Heising-Simons Foundation under Grant 2024-4848.

\addcontentsline{toc}{section}{References}
\bibliographystyle{JHEP}
\bibliography{ref2}

@article{Maldacena:2026jqd,
    author = "Maldacena, Juan and Maloney, Alexander and McPeak, Brian",
    title = "{Wormholes and the imaginary distance bound}",
    eprint = "2605.05336",
    archivePrefix = "arXiv",
    primaryClass = "hep-th",
    month = "5",
    year = "2026"
}

@article{Halliwell:1989pu,
    author = "Halliwell, Jonathan J. and Myers, Robert C.",
    title = "{Multiple Sphere Configurations in the Path Integral Representation of the Wave Function of the Universe}",
    reportNumber = "NSF-ITP-89-124",
    doi = "10.1103/PhysRevD.40.4011",
    journal = "Phys. Rev. D",
    volume = "40",
    pages = "4011",
    year = "1989"
}

@article{Garcia-Garcia:2020ttf,
    author = "Garc\'\i{}a-Garc\'\i{}a, Antonio M. and Godet, Victor",
    title = "{Euclidean wormhole in the Sachdev-Ye-Kitaev model}",
    eprint = "2010.11633",
    archivePrefix = "arXiv",
    primaryClass = "hep-th",
    doi = "10.1103/PhysRevD.103.046014",
    journal = "Phys. Rev. D",
    volume = "103",
    number = "4",
    pages = "046014",
    year = "2021"
}

@article{H,
author = "C. J. Howls",
title = "Hyperasymptotics For Multidimensional Integrals, Exact Remainder Terms And The Global Connection
Problem",
journal = "Proc. R. Soc. London",
volume = "453",
year = "1997",
pages =  "2271"
}

@article{Marolf:2021kjc,
    author = "Marolf, Donald and Santos, Jorge E.",
    title = "{AdS Euclidean wormholes}",
    eprint = "2101.08875",
    archivePrefix = "arXiv",
    primaryClass = "hep-th",
    doi = "10.1088/1361-6382/ac2cb7",
    journal = "Class. Quant. Grav.",
    volume = "38",
    number = "22",
    pages = "224002",
    year = "2021"
}

@article{Arkani-Hamed:2007cpn,
    author = "Arkani-Hamed, Nima and Orgera, Jacopo and Polchinski, Joseph",
    title = "{Euclidean wormholes in string theory}",
    eprint = "0705.2768",
    archivePrefix = "arXiv",
    primaryClass = "hep-th",
    doi = "10.1088/1126-6708/2007/12/018",
    journal = "JHEP",
    volume = "12",
    pages = "018",
    year = "2007"
}

@article{Louko:1995jw,
    author = "Louko, Jorma and Sorkin, Rafael D.",
    title = "{Complex actions in two-dimensional topology change}",
    eprint = "gr-qc/9511023",
    archivePrefix = "arXiv",
    reportNumber = "SU-GP-95-5-1, WISC-MILW-95-TH-16, MDDP-PP-96-40",
    doi = "10.1088/0264-9381/14/1/018",
    journal = "Class. Quant. Grav.",
    volume = "14",
    pages = "179--204",
    year = "1997"
}

@article{Colin-Ellerin:2020mva,
    author = "Colin-Ellerin, Sean and Dong, Xi and Marolf, Donald and Rangamani, Mukund and Wang, Zhencheng",
    title = "{Real-time gravitational replicas: Formalism and a variational principle}",
    eprint = "2012.00828",
    archivePrefix = "arXiv",
    primaryClass = "hep-th",
    doi = "10.1007/JHEP05(2021)117",
    journal = "JHEP",
    volume = "05",
    pages = "117",
    year = "2021"
}

@article{AxionWormholes,
    author = "Held, Jesse and Kaplan, Molly and Marolf, Donald and Wang, Zhencheng",
    title = "{Axion Wormholes and the AdS/CFT Factorization Problem}",
    eprint = "2601.02507",
    archivePrefix = "arXiv",
    primaryClass = "hep-th",
    month = "1",
    year = "2026"
}

@article{Maldacena:2004rf,
    author = "Maldacena, Juan Martin and Maoz, Liat",
    title = "{Wormholes in AdS}",
    eprint = "hep-th/0401024",
    archivePrefix = "arXiv",
    reportNumber = "ITFA-2003-57",
    doi = "10.1088/1126-6708/2004/02/053",
    journal = "JHEP",
    volume = "02",
    pages = "053",
    year = "2004"
}

@article{Saad:2018bqo,
    author = "Saad, Phil and Shenker, Stephen H. and Stanford, Douglas",
    title = "{A semiclassical ramp in SYK and in gravity}",
    eprint = "1806.06840",
    archivePrefix = "arXiv",
    primaryClass = "hep-th",
    month = "6",
    year = "2018"
}

@article{Horowitz:2025zpx,
    author = "Horowitz, Gary T. and Marolf, Donald and Santos, Jorge E.",
    title = "{Constraints are not enough}",
    eprint = "2505.13600",
    archivePrefix = "arXiv",
    primaryClass = "hep-th",
    month = "5",
    year = "2025"
}

@article{Friedman:1993ty,
    author = "Friedman, John L. and Schleich, Kristin and Witt, Donald M.",
    title = "{Topological censorship}",
    eprint = "gr-qc/9305017",
    archivePrefix = "arXiv",
    reportNumber = "PRINT-93-0448 (SANTA-BARBARA,ITP), NSF-ITP-93-80",
    doi = "10.1103/PhysRevLett.71.1486",
    journal = "Phys. Rev. Lett.",
    volume = "71",
    pages = "1486--1489",
    year = "1993",
    note = "[Erratum: Phys.Rev.Lett. 75, 1872 (1995)]"
}

@article{Held:2024qcl,
    author = "Held, Jesse and Liu, Xiaoyi and Marolf, Donald and Wang, Zhencheng",
    title = "{Euclidean and complex geometries from real-time computations of gravitational R{\'e}nyi
    entropies}",
    eprint = "2409.17428",
    archivePrefix = "arXiv",
    primaryClass = "hep-th",
    doi = "10.1007/JHEP02(2025)136",
    journal = "JHEP",
    volume = "02",
    pages = "136",
    year = "2025"
}

@article{Gibbons:1978ac,
    author = "Gibbons, G. W. and Hawking, S. W. and Perry, M. J.",
    title = "{Path Integrals and the Indefiniteness of the Gravitational Action}",
    reportNumber = "PRINT-78-0375 (CAMBRIDGE)",
    doi = "10.1016/0550-3213(78)90161-X",
    journal = "Nucl. Phys. B",
    volume = "138",
    pages = "141--150",
    year = "1978"
}

@article{Dong:2016hjy,
    author = "Dong, Xi and Lewkowycz, Aitor and Rangamani, Mukund",
    title = "{Deriving covariant holographic entanglement}",
    eprint = "1607.07506",
    archivePrefix = "arXiv",
    primaryClass = "hep-th",
    doi = "10.1007/JHEP11(2016)028",
    journal = "JHEP",
    volume = "11",
    pages = "028",
    year = "2016"
}

@article{Marolf:Thermodynamics,
    author = "Marolf, Donald",
    title = "{Gravitational thermodynamics without the conformal factor problem: partition functions and
    Euclidean saddles from Lorentzian path integrals}",
    eprint = "2203.07421",
    archivePrefix = "arXiv",
    primaryClass = "hep-th",
    doi = "10.1007/JHEP07(2022)108",
    journal = "JHEP",
    volume = "07",
    pages = "108",
    year = "2022"
}

@article{Hertog:Axion-Saxion_Wormholes_Stable,
    author = "Hertog, T. and Maenaut, S. and Missoni, B. and Tielemans, R. and Van Riet, T.",
    title = "{Stability of axion-saxion wormholes}",
    eprint = "2405.02072",
    archivePrefix = "arXiv",
    primaryClass = "hep-th",
    doi = "10.1007/JHEP11(2024)151",
    journal = "JHEP",
    volume = "11",
    pages = "151",
    year = "2024"
}

@article{Giddings:Original_Axion_Wormhole,
    author = "Giddings, Steven B. and Strominger, Andrew",
    title = "{Axion Induced Topology Change in Quantum Gravity and String Theory}",
    reportNumber = "HUTP-87-A067",
    doi = "10.1016/0550-3213(88)90446-4",
    journal = "Nucl. Phys. B",
    volume = "306",
    pages = "890--907",
    year = "1988"
}

@article{loveridge:Axion_Wormhole_Alternate_Topologies,
    author = "Loveridge, Andrew and Sun, Hao-Yu",
    title = "{AdS3 axion wormholes as stable contributions to the Euclidean gravitational path integral}",
    eprint = "2504.10868",
    archivePrefix = "arXiv",
    primaryClass = "hep-th",
    month = "4",
    year = "2025"
}

@article{Chandra:2022bqq,
    author = "Chandra, Jeevan and Collier, Scott and Hartman, Thomas and Maloney, Alexander",
    title = "{Semiclassical 3D gravity as an average of large-c CFTs}",
    eprint = "2203.06511",
    archivePrefix = "arXiv",
    primaryClass = "hep-th",
    doi = "10.1007/JHEP12(2022)069",
    journal = "JHEP",
    volume = "12",
    pages = "069",
    year = "2022"
}

@article{Held:2026bbo,
    author = "Held, Jesse and Kaplan, Molly and Marolf, Donald and Wang, Zhencheng",
    title = "{Lorentzian path integrals and Jackiw-Teitelboim wormholes with imaginary scalars}",
    eprint = "2601.09932",
    archivePrefix = "arXiv",
    primaryClass = "hep-th",
    doi = "10.1007/JHEP07(2026)215",
    journal = "JHEP",
    volume = "07",
    pages = "215",
    year = "2026"
}

@article{Balasubramanian:2022gmo,
    author = "Balasubramanian, Vijay and Lawrence, Albion and Magan, Javier M. and Sasieta, Martin",
    title = "{Microscopic Origin of the Entropy of Black Holes in General Relativity}",
    eprint = "2212.02447",
    archivePrefix = "arXiv",
    primaryClass = "hep-th",
    doi = "10.1103/PhysRevX.14.011024",
    journal = "Phys. Rev. X",
    volume = "14",
    number = "1",
    pages = "011024",
    year = "2024"
}

@inproceedings{Hartle:2020glw,
    title = "{The Conformal Rotation in Linearised Gravity}",
    author = "Hartle, James B. and Schleich, Kristen",
    booktitle="Quantum Field Theory and Quantum Statistics",
    editor="I. A. Batalin, C.
J. Isham and G. A. Vilkovisky",
published="Adam Hilger",
pages = "{67-87}",
    eprint = "2004.06635",
    archivePrefix = "arXiv",
    primaryClass = "gr-qc",
    month = "4",
    year = "1987"
}

@article{Hawking:1996,
    author = "Hawking, S. W. and Horowitz, Gary T.",
    title = "{The Gravitational Hamiltonian, action, entropy and surface terms}",
    eprint = "gr-qc/9501014",
    archivePrefix = "arXiv",
    reportNumber = "DAMTP-R-94-52, UCSBTH-94-37",
    doi = "10.1088/0264-9381/13/6/017",
    journal = "Class. Quant. Grav.",
    volume = "13",
    pages = "1487--1498",
    year = "1996"
}

@article{Schleich:1987fm,
    author = "Schleich, K.",
    editor = "Gibbons, G. W. and Hawking, S. W.",
    title = "{Conformal Rotation in Perturbative Gravity}",
    doi = "10.1103/PhysRevD.36.2342",
    journal = "Phys. Rev. D",
    volume = "36",
    pages = "2342--2363",
    year = "1987"
}

@article{Mazur:1989by,
    author = "Mazur, Pawel O. and Mottola, Emil",
    title = "{The Gravitational Measure, Solution of the Conformal Factor Problem and Stability of the Ground
    State of Quantum Gravity}",
    reportNumber = "UFIFT-AST-89-3, LA-UR-89-340",
    doi = "10.1016/0550-3213(90)90268-I",
    journal = "Nucl. Phys. B",
    volume = "341",
    pages = "187--212",
    year = "1990"
}

@article{Dasgupta:2001ue,
    author = "Dasgupta, A. and Loll, R.",
    title = "{A Proper time cure for the conformal sickness in quantum gravity}",
    eprint = "hep-th/0103186",
    archivePrefix = "arXiv",
    reportNumber = "AEI-2001-020",
    doi = "10.1016/S0550-3213(01)00227-9",
    journal = "Nucl. Phys. B",
    volume = "606",
    pages = "357--379",
    year = "2001"
}

@article{Marolf:2020rpm,
    author = "Marolf, Donald and Maxfield, Henry",
    title = "{Observations of Hawking radiation: the Page curve and baby universes}",
    eprint = "2010.06602",
    archivePrefix = "arXiv",
    primaryClass = "hep-th",
    doi = "10.1007/JHEP04(2021)272",
    journal = "JHEP",
    volume = "04",
    pages = "272",
    year = "2021"
}

@article{Ambjorn:2002gr,
    author = "Ambjorn, J. and Dasgupta, A. and Jurkiewicz, J. and Loll, R.",
    title = "{A Lorentzian cure for Euclidean troubles}",
    eprint = "hep-th/0201104",
    archivePrefix = "arXiv",
    reportNumber = "SPIN-2001-33",
    doi = "10.1016/S0920-5632(01)01903-X",
    journal = "Nucl. Phys. B Proc. Suppl.",
    volume = "106",
    pages = "977--979",
    year = "2002"
}

@article{Marolf:2025evo,
    author = "Marolf, Donald and Missoni, Bruno",
    title = "{Euclidean wormholes stability analysis revisited}",
    eprint = "2505.21118",
    archivePrefix = "arXiv",
    primaryClass = "hep-th",
    month = "5",
    year = "2025"
}

@article{Colin-Ellerin:2021jev,
    author = "Colin-Ellerin, Sean and Dong, Xi and Marolf, Donald and Rangamani, Mukund and Wang, Zhencheng",
    title = "{Real-time gravitational replicas: low dimensional examples}",
    eprint = "2105.07002",
    archivePrefix = "arXiv",
    primaryClass = "hep-th",
    doi = "10.1007/JHEP08(2021)171",
    journal = "JHEP",
    volume = "08",
    pages = "171",
    year = "2021"
}

@inproceedings{Giddings:1990yj,
    author = "Giddings, Steven B.",
    title = "{Wormholes, the conformal factor, and the cosmological constant}",
    booktitle = "{International Colloquium on Modern Quantum Field Theory}",
    reportNumber = "HUTP-90-A010",
    month = "5",
    year = "1990"
}

@article{Giddings:1989ny,
    author = "Giddings, Steven B.",
    title = "{The Conformal Factor and the Cosmological Constant}",
    reportNumber = "HUTP-89/A056",
    doi = "10.1142/S0217751X9000163X",
    journal = "Int. J. Mod. Phys. A",
    volume = "5",
    pages = "3811--3830",
    year = "1990"
}

@article{Marolf:1996gb,
    author = "Marolf, Donald",
    title = "{Path integrals and instantons in quantum gravity: Minisuperspace models}",
    eprint = "gr-qc/9602019",
    archivePrefix = "arXiv",
    reportNumber = "UCSBTH-96-01",
    doi = "10.1103/PhysRevD.53.6979",
    journal = "Phys. Rev. D",
    volume = "53",
    pages = "6979--6990",
    year = "1996"
}

@article{Feldbrugge:2017kzv,
    author = "Feldbrugge, Job and Lehners, Jean-Luc and Turok, Neil",
    title = "{Lorentzian Quantum Cosmology}",
    eprint = "1703.02076",
    archivePrefix = "arXiv",
    primaryClass = "hep-th",
    doi = "10.1103/PhysRevD.95.103508",
    journal = "Phys. Rev. D",
    volume = "95",
    number = "10",
    pages = "103508",
    year = "2017"
}

@article{Gratton:1999ya,
    author = "Gratton, Steven and Turok, Neil",
    title = "{Cosmological perturbations from the no boundary Euclidean path integral}",
    eprint = "astro-ph/9902265",
    archivePrefix = "arXiv",
    doi = "10.1103/PhysRevD.60.123507",
    journal = "Phys. Rev. D",
    volume = "60",
    pages = "123507",
    year = "1999"
}

@article{Feldbrugge:2017fcc,
    author = "Feldbrugge, Job and Lehners, Jean-Luc and Turok, Neil",
    title = "{No smooth beginning for spacetime}",
    eprint = "1705.00192",
    archivePrefix = "arXiv",
    primaryClass = "hep-th",
    doi = "10.1103/PhysRevLett.119.171301",
    journal = "Phys. Rev. Lett.",
    volume = "119",
    number = "17",
    pages = "171301",
    year = "2017"
}

@article{Feldbrugge:2017mbc,
    author = "Feldbrugge, Job and Lehners, Jean-Luc and Turok, Neil",
    title = "{No rescue for the no boundary proposal: Pointers to the future of quantum cosmology}",
    eprint = "1708.05104",
    archivePrefix = "arXiv",
    primaryClass = "hep-th",
    doi = "10.1103/PhysRevD.97.023509",
    journal = "Phys. Rev. D",
    volume = "97",
    number = "2",
    pages = "023509",
    year = "2018"
}

@article{andriolo:Axion_wormholes_massive_dilaton,
    author = "Andriolo, Stefano and Shiu, Gary and Soler, Pablo and Van Riet, Thomas",
    title = "{Axion wormholes with massive dilaton}",
    eprint = "2205.01119",
    archivePrefix = "arXiv",
    primaryClass = "hep-th",
    reportNumber = "CTPU-PTC-22-07",
    doi = "10.1088/1361-6382/ac8fdc",
    journal = "Class. Quant. Grav.",
    volume = "39",
    number = "21",
    pages = "215014",
    year = "2022"
}

@article{Witten:2026twr,
    author = "Witten, Edward",
    title = "{Duality and Axion Wormholes}",
    eprint = "2601.01587",
    archivePrefix = "arXiv",
    primaryClass = "hep-th",
    month = "1",
    year = "2026"
}

@article{Carlip:1993sa,
    author = "Carlip, Steven and Teitelboim, Claudio",
    title = "{The Off-shell black hole}",
    eprint = "gr-qc/9312002",
    archivePrefix = "arXiv",
    reportNumber = "IASSNS-HEP-93-84, UCD-93-34",
    doi = "10.1088/0264-9381/12/7/011",
    journal = "Class. Quant. Grav.",
    volume = "12",
    pages = "1699--1704",
    year = "1995"
}

@article{Cotler:2021cqa,
    author = "Cotler, Jordan and Jensen, Kristan",
    title = "{Wormholes and black hole microstates in AdS/CFT}",
    eprint = "2104.00601",
    archivePrefix = "arXiv",
    primaryClass = "hep-th",
    doi = "10.1007/JHEP09(2021)001",
    journal = "JHEP",
    volume = "09",
    pages = "001",
    year = "2021"
}

@article{aguilar:Axion_dS_Wormholes,
    author = "Aguilar-Gutierrez, Sergio E. and Hertog, Thomas and Tielemans, Rob and van der Schaar, Jan Pieter
    and Van Riet, Thomas",
    title = "{Axion-de Sitter wormholes}",
    eprint = "2306.13951",
    archivePrefix = "arXiv",
    primaryClass = "hep-th",
    doi = "10.1007/JHEP11(2023)225",
    journal = "JHEP",
    volume = "11",
    pages = "225",
    year = "2023"
}

@article{andrade:BeyondUnitarityBound,
    author = "Andrade, Tomas and Marolf, Donald",
    title = "{AdS/CFT beyond the unitarity bound}",
    eprint = "1105.6337",
    archivePrefix = "arXiv",
    primaryClass = "hep-th",
    doi = "10.1007/JHEP01(2012)049",
    journal = "JHEP",
    volume = "01",
    pages = "049",
    year = "2012"
}

@unpublished{AKM,
  author    = {Antonini, Stefano and Kolanowski, Maciej and Marolf, Donald},
  title     = {to appear},
  note      = {To appear },
  year      = {2026}
}

@article{Loges:2023ypl,
    author = "Loges, Gregory J. and Shiu, Gary and Van Riet, Thomas",
    title = "{A 10d construction of Euclidean axion wormholes in flat and AdS space}",
    eprint = "2302.03688",
    archivePrefix = "arXiv",
    primaryClass = "hep-th",
    reportNumber = "KEK-TH-2495",
    doi = "10.1007/JHEP06(2023)079",
    journal = "JHEP",
    volume = "06",
    pages = "079",
    year = "2023"
}

@article{Loges:2022nuw,
    author = "Loges, Gregory J. and Shiu, Gary and Sudhir, Nidhi",
    title = "{Complex saddles and Euclidean wormholes in the Lorentzian path integral}",
    eprint = "2203.01956",
    archivePrefix = "arXiv",
    primaryClass = "hep-th",
    doi = "10.1007/JHEP08(2022)064",
    journal = "JHEP",
    volume = "08",
    pages = "064",
    year = "2022"
}

@article{Brown:2017wpl,
    author = "Brown, Jon and Cole, Alex and Shiu, Gary and Cottrell, William",
    title = "{Gravitational decoupling and the Picard-Lefschetz approach}",
    eprint = "1710.04737",
    archivePrefix = "arXiv",
    primaryClass = "hep-th",
    reportNumber = "MAD-TH-17-06",
    doi = "10.1103/PhysRevD.97.025002",
    journal = "Phys. Rev. D",
    volume = "97",
    number = "2",
    pages = "025002",
    year = "2018"
}

@article{Garriga:1997wz,
    author = "Garriga, Jaume and Montes, Xavier and Sasaki, Misao and Tanaka, Takahiro",
    title = "{Canonical quantization of cosmological perturbations in the one-bubble open universe}",
    eprint = "astro-ph/9706229",
    archivePrefix = "arXiv",
    reportNumber = "UAB-FT-418, OU-TAP-63",
    doi = "10.1016/S0550-3213(97)00780-3",
    journal = "Nucl. Phys. B",
    volume = "513",
    pages = "343--374",
    year = "1998",
    note = "[Erratum: Nucl.Phys.B 551, 511--511 (1999)]"
}

@article{Kol:2006ga,
    author = "Kol, Barak",
    title = "{The Power of Action: The Derivation of the Black Hole Negative Mode}",
    eprint = "hep-th/0608001",
    archivePrefix = "arXiv",
    doi = "10.1103/PhysRevD.77.044039",
    journal = "Phys. Rev. D",
    volume = "77",
    pages = "044039",
    year = "2008"
}

@article{Cotler_2021:ConstrainedInstantons,
   title={Gravitational constrained instantons},
    eprint = "2010.02241",
    archivePrefix = "arXiv",
    primaryClass = "hep-th",
   volume={104},
   pages="L081501",
   ISSN={2470-0029},
   url={http://dx.doi.org/10.1103/PhysRevD.104.L081501},
   DOI={10.1103/physrevd.104.l081501},
   number={8},
   journal={Physical Review D},
   publisher={American Physical Society (APS)},
   author={Cotler, Jordan and Jensen, Kristan},
   year={2021},
   month=oct }

@article{Garcia-Garcia:2022zmo,
    author = "Garc{\'\i}a-Garc{\'\i}a, Antonio M. and Godet, Victor and Yin, Can and Zheng, Jie Ping",
    title = "{Euclidean-to-Lorentzian wormhole transition and gravitational symmetry breaking in the
    Sachdev-Ye-Kitaev model}",
    eprint = "2204.08558",
    archivePrefix = "arXiv",
    primaryClass = "hep-th",
    doi = "10.1103/PhysRevD.106.046008",
    journal = "Phys. Rev. D",
    volume = "106",
    number = "4",
    pages = "046008",
    year = "2022"
}

@article{Wang:2025bcx,
    author = "Wang, Diandian and Wang, Zhencheng and Wei, Zixia",
    title = "{Wormholes with ends of the world}",
    eprint = "2504.12278",
    archivePrefix = "arXiv",
    primaryClass = "hep-th",
    doi = "10.1007/JHEP09(2025)166",
    journal = "JHEP",
    volume = "09",
    pages = "166",
    year = "2025"
}
\end{document}